\documentclass[reprint,pra,groupedaddress,showpacs,twocolumn]{revtex4}
\usepackage{units}
\usepackage{amsmath}
\usepackage{amssymb}
\usepackage{graphicx}
\usepackage{bm}

\newcommand{\gper}{\gamma_\perp}
\newcommand{\gpar}{\gamma_\parallel}
\newcommand{\be}{\begin{equation}}
\newcommand{\ee}{\end{equation}}
\newcommand{\bea}{\begin{eqnarray}}
\newcommand{\eea}{\end{eqnarray}}

\begin{document}
\title{Steady-state Ab Initio Laser Theory: Generalizations and Analytic Results}

\author{Li Ge, Y.~D.~Chong, and A.~Douglas Stone}
\affiliation{Department of Applied Physics, P. O. Box 208284, Yale University,
New Haven, CT 06520-8284, USA}

\date{\today}

\begin{abstract}
We improve the steady-state ab initio laser theory (SALT) of T\"ureci {\it et al}.~by expressing its fundamental self-consistent equation in a basis set of threshold constant flux states that contains the exact threshold lasing mode.  For cavities with non-uniform index and/or non-uniform gain, the new basis set allows the steady-state lasing properties to be computed with much greater efficiency.  This formulation of the SALT can be solved in the single-pole approximation, which gives the intensities and thresholds, including the effects of nonlinear hole-burning interactions to all orders, with negligible computational effort.  The approximation yields a number of analytic predictions, including a ``gain-clamping'' transition at which strong modal interactions suppress all higher modes.  We show that the single-pole approximation agrees well with exact SALT calculations, particularly for high-Q cavities.  Within this range of validity, it provides an extraordinarily efficient technique for modeling realistic and complex lasers.
\end{abstract}

\pacs{42.55.Ah,42.60.Da,42.55.Zz,02.70.Hm}
\maketitle

\section{Introduction}
The foundation of our understanding of lasers is semiclassical laser theory, in which the gain medium is treated quantum-mechanically and the electromagnetic fields are treated classically.  The pioneering work of Haken \cite{Haken} and Lamb \cite{Lamb} showed that the Maxwell-Bloch (MB) equations, in which the gain medium is modeled by an ensemble of two-level atoms, successfully describe the principal properties of lasers, including modal thresholds, lasing frequencies, output power, the structure of the electromagnetic fields inside and outside of the laser cavity, as well as dynamical effects such as relaxation oscillations and mode, phase and frequency locking.  The only properties that cannot be obtained from the semiclassical theory are those that depend on quantum fluctuations of the electromagnetic field, such as the laser linewidth, amplified spontaneous emission and photon statistics.

Because the MB equations are coupled nonlinear equations in space and time, few purely analytic results could be obtained from the theory.  Those obtained generally relied on a number of drastic approximations: the mode structure was assumed to be simple (e.g.~spatially uniform or low-order gaussian modes), the openness of the laser system was handled either through adding phenomenological damping to closed cavity modes or by approximating the lasing modes as quasimodes of the passive cavity, and the nonlinear modal interactions were either ignored or simplified by solving the equations near threshold.  Where such approximations could not be employed, reliable theoretical results could only be obtained from brute force time-domain simulations of the MB equations \cite{Andreasen}, or their multi-level generalizations.

The last two decades have seen the emergence of novel laser systems based on complex resonators, driven by advances in microfabrication and motivated by applications to integrated on-chip optics, as well as by basic scientific interest.  Examples are VCSELs \cite{vcsel}, microdisk \cite{Levi_ieee92,Chang_book96}, spiral \cite{Chern_apl03} and (wave-chaotic) deformed disk lasers \cite{Nockel_nat97,Gmachl_sci98,Chang_JOSAB00,Wiersig_prl08}, photonic crystal lasers \cite{Painter_sci1999,Noda_nat00}, and random lasers \cite{Cao_prl99, Mujumdar_prl04}.  The analytical theory existing at the time was not readily applicable to these complex systems; the random laser, in particular, poses a difficult conceptual challenge as the corresponding laser ``cavity'' has finesse smaller than unity, meaning that it has no isolated passive cavity resonances.  At the same time, the complexity of some of these structures pushed realistic simulations of the lasing equations to the limits of practicality.  Thus, there was a need for a robust semiclassical laser theory incorporating a more accurate treatment of the cavity modes, including both spatial complexity and openness, as well as the effects of nonlinear modal interactions.

Such a theory has been proposed by T\"ureci and Stone \cite{TS}.  These authors employed one key approximation, originally introduced by Haken \cite{Haken,TSG OE,Oleg}: the inversion is assumed to be time-independent, implying the absence of definite phase relationships between the lasing modes (hence ignoring phenomena such as mode and phase locking).
This stationary inversion approximation (SIA), also called the ``free-running" approximation \cite{sslbook,mandelbook}, had been previously employed in combination with a third-order treatment of the non-linearity, to arrive at the Haken-Sauermann (HS) equations of multimode laser theory \cite{Haken,HS}.  About a decade ago, Mandel and coworkers \cite{Mandel1,Mandel2} also used the SIA and went beyond the HS equations to give an infinite-order treatment of non-linear interactions in the Fabry-Perot cavity with non-uniform pumping.  Both the HS and Mandel approaches neglected the effect of the openness of the cavity on the lasing modes (we compare our theory to the Mandel approach in Appendix B, having already compared to HS in Ref. \cite{TS}).  In contrast the current approach is formulated for arbitrary cavities and pump profile and treats the
openness exactly.  By seeking only the steady-state solutions of the MB equations, T\"ureci and Stone \cite{TS} derived a set of self-consistent time-independent nonlinear equations for the lasing modes and frequencies at a given pump.  Subsequently, T\"ureci, Stone and Ge \cite{TSG PRA} and Ge \textit{et~al.}  \cite{TSG OE} developed an iterative algorithm for solving these nonlinear equations, eliminated the slowly-varying envelope approximation, and confirmed that the resulting solutions agreed to high accuracy with the steady-state results of time-domain simulations of the MB equations.  In 2008, T\"ureci \textit{et~al.} \cite{TSG Science, TSG Nonlinearity} showed that even multimode random lasing in two dimensions (2D) could be efficiently calculated using this method.  We will refer to this approach as Steady-state Ab Initio Laser Theory (SALT) \cite{AISC}.  The term ``ab initio" refers to the fact that the only inputs are the dielectric function for the passive cavity and a few parameters to describe the gain medium.  The SALT method thus bridges the gap between oversimplified analytical approaches and time-domain simulations.  Unlike the former, it describes laser cavities of arbitrary complexity and openness, making no assumptions about the nature of the lasing modes or frequencies, or the proximity to threshold.  Unlike the latter, it yields direct semi-analytic insights into the lasing solutions.  Furthermore, the SALT method is in general much more computationally efficient than time-domain simulations; due to the elimination of the time variable, it allows for calculations that are impractical in the time domain due to limitations in computer speed or memory.

In the present work, we present a significant improvement to the SALT method by introducing a new basis set that always contains the exact threshold lasing solution.  The properties of the new basis functions allow us to compute the lasing solutions above threshold more efficiently than before.  They also allow us to derive an approximation to the full SALT for high-Q lasing cavities, which we abbreviate as
SPA-SALT (single-pole-approximation SALT) which is valid well above threshold in contrast to the HS theory \cite{HS}. The SPA-SALT approximation yields solutions with negligible computational effort, once the threshold lasing properties are known.  From this simplified theory, we derive several analytic results for the lasing behavior above threshold, including relatively simple formulas for the thresholds of higher lasing modes.  These results hold to infinite order in the nonlinear modal interactions and are hence quantitatively reliable.  Strikingly, these results predict a ``gain-clamping'' transition, in which higher modes are prevented from turning on at any pump, despite the non-uniformity of the lasing modes.

The remainder of this paper is organized as follows. In Sec.~\ref{Ab-initio laser equations}, we review the previous formulation of the SALT and the solution method based on the basis set of constant-flux states, and describes the limitations imposed by this basis set. In Sec.~\ref{Basis functions section}, we present the new basis set and the formulation of the SALT in terms of this basis, and examine the efficiency of the new solution method.  In Sec.~\ref{Approximate analytic solution of SALT equations}, we derive the simplified form of the SALT equations arising from the ``single-pole approximation'' (SPA-SALT).  We then solve these equations analytically, and demonstrate good agreement with the exact SALT solutions. In the appendices we derive the power output equations of the SALT, compare the SPA-SALT to the earlier Mandel approach \cite{Mandel1,Mandel2}, and calculate perturbative corrections to the SPA-SALT.

\section{Ab-initio laser equations}
\label{Ab-initio laser equations}

\subsection{The SALT Equations}

The SALT description of lasing begins with the MB equations for an ensemble of two level atoms interacting with a classical electromagnetic field:
\begin{eqnarray}
  &&\nabla^2 E^+ - \epsilon_c(\vec{r})\, \ddot{E}^+ = 4\pi \ddot{P}^+, \label{MB1}\\
  &&\dot{P}^+ = -i(k_a-i\gamma_\perp) P^+ + \frac{g^2}{i\hbar} E^+ D, \label{MB2}\\
  &&\dot{D} = \gamma_\parallel (D_0 - D) - \frac{2}{i\hbar}
  \left[E^+(P^+)^* - P^+(E^+)^*\right]. \label{MB3}
\end{eqnarray}
Here, we restrict the fields to one dimension (1D), or to the transverse magnetic (TM) polarization in 2D, so that the electric and polarization fields are scalars (the generalization to TE modes in 2D is straightforward). Their positive-frequency components are $E^+(\vec{r},t)$ and $P^+(\vec{r},t)$; in these equations, we have made use of the rotating-wave approximation (RWA). Note that we have not used the standard slowly-varying envelope approximation, employed in most treatments to eliminate second time derivatives; this approximation gives no benefit in the SALT approach, and is unnecessary \cite{TSG OE}. We have taken the speed of light in vacuum $c$ to be unity; wavevector and frequency will be distinguished by the context. $D(\vec{r},t)$ is the population inversion, and $D_0(\vec{r})$ is the pump; $k_a$ is the frequency of the gain center, $\gamma_\perp$ is the gain width (polarization dephasing rate), $\gamma_\parallel$ is the population relaxation rate, $g$ is the dipole matrix element, and $\epsilon_c(\vec{r})$ is the cavity dielectric function, which in general is complex and includes the material absorption inside the cavity. Arbitrary cavity elements, such as mirrors can be represented by an appropriate choice of $\epsilon_c(\vec{r})$, although we will focus on dielectric cavities in our examples below. We assume that the $E^+$ and $P^+$ fields obey a multi-mode ansatz
\begin{align}
  \begin{aligned}
    E^+(\vec{r},t) &= \sum_{\mu=1}^N \Psi_\mu(\vec{r})\,e^{-ik_\mu t}, \\
    P^+(\vec{r},t) &= \sum_{\mu=1}^N p_\mu(\vec{r})\,e^{-ik_\mu t},
    \label{mode ansatz}
  \end{aligned}
\end{align}
where the indices $\mu = 1,2,\cdots,N$ label the different lasing modes.  The total number of modes, $N$, is not given, but increases in unit steps from zero as we increase the pump strength $D_0$.  The values of $D_0$ at which each step occurs are the (interacting) modal thresholds, to be determined self-consistently from the theory. The real numbers $k_\mu$ are the lasing frequencies of the modes, which will also be determined self-consistently.

We insert the ansatz (\ref{mode ansatz}) into Eqs.~(\ref{MB1}--\ref{MB3}), and employ the stationary inversion approximation $\dot{D} = 0$.  The result is a set of coupled nonlinear differential equations, which are the fundamental equations of the SALT \cite{LiThesis}:
\begin{eqnarray}
  \left[\nabla^2 + \left(\epsilon_c(\vec{r}) + \frac{\gper D(\vec{r})}{k_\mu - k_a + i\gamma_\perp} \right)k_\mu^2\right]
  \Psi_\mu(\vec{r}) = 0, \label{TSG1} \\
  D(\vec{r}) = D_0(\vec{r}) \, \left[1 +\sum_{\nu=1}^N \Gamma_\nu |\Psi_\nu(\vec{r})|^2\right]^{-1}. \label{TSG2}
\end{eqnarray}
$\Psi$ and $D$ are now dimensionless, measured in their natural units $e_c = \hbar\sqrt{\gpar\gper}/(2g)$ and $d_c = \hbar\gper/(4\pi g^2)$, and $\Gamma_\nu \equiv \gper^2/(\gper^2 + (k_\nu-k_a)^2)$ is the Lorentzian gain curve evaluated at frequency $k_\nu$. Eq.~(\ref{TSG1}) is simply a wave equation for the electric field mode $\Psi_\mu$, with an effective dielectric function consisting of both the ``passive'' contribution $\epsilon_c(\vec{r})$ and an ``active'' contribution
from the gain medium.  The latter is frequency-dependent, and has both a real part and a negative (amplifying) imaginary part.  It also includes
infinite-order nonlinear ``hole-burning'' modal interactions, seen in the $|\Psi_\nu|^2$ dependence of (\ref{TSG2}).  In addition, we make the key
requirement that $\Psi_\mu$ must be purely out-going outside the cavity; it is this condition that makes the problem non-Hermitian.  It is worth noting that
the stationary inversion approximation is not needed until at least two modes are above threshold, so (\ref{TSG2}) is exact for single-mode lasing up to and including the second threshold (aside from the well-obeyed RWA).

Let us define a finite cavity region $C$, such that
\begin{equation}
  D_0(\vec{r}) = 0 \;\;\;\textrm{and}\;\;\; \epsilon_c(\vec{r}) = n_0^2, \quad
  \vec{r} \notin C. \label{eq:boundary}
\end{equation}
Although we call $C$ the ``cavity'' region, $\epsilon_c(\vec{r})$ need not be discontinuous at its boundary.  The theory applies, for instance, to random lasers lacking any well-defined boundary \cite{TSG   Science,TSG Nonlinearity}.  For our purposes, $C$ simply defines a surface of last scattering (or last amplification), a region outside of which there is no dielectric nor gain material to affect the free propagation of waves.

We write the external pump as
\begin{equation}
  D_0(\vec{r}) = D_0 \; F(\vec{r}), \quad \vec{r} \in C,\label{eq:F(r)}
\end{equation}
where $D_0$ is the ``pump strength'' and $F(\vec{r})$ a fixed ``pump profile'', both real quantities.  The simplest case, $F(\vec{r}) = 1$, corresponds to uniform pumping within the cavity.  In general $F(\vec{r})$ need not be uniform, e.g.~if the pump is a finite laser spot or the gain material is distributed unevenly.

The lasing equation now becomes
\begin{equation}
  \left[\nabla^2 + \left(\epsilon_c(\vec{r}) + \frac{\gamma_\mu D_0 F(\vec{r})}{1+h(\vec{r})}
    \right)k_\mu^2\right] \Psi_\mu(\vec{r}) = 0 \label{TSG1a}
\end{equation}
in which $h(\vec{r}) \equiv \sum_\nu \Gamma_\nu|\Psi(\vec{r})|^2$ represents the spatial hole burning effect.
Here we have introduced the abbreviation
\begin{equation}
  \gamma_\mu \equiv \gamma_\perp/(k_\mu - k_a + i \gamma_\perp).
\end{equation}
Previous treatments of the SALT \cite{TS,TSG OE,TSG Science,TSG Nonlinearity} proceeded by inverting (\ref{TSG1a}) via the Green's function to yield an equivalent integral equation, but for our purposes it is more convenient to retain the differential form.

\subsection{Modal Output Power}
Using Eq.~(\ref{TSG1a}) we can determine the unknown lasing frequencies $k_\mu$ and mode fields $\Psi_\mu(\vec{r})$.  From these quantities, all other properties associated with the semiclassical steady state can be derived. For instance, an important quantity not treated explicitly in earlier versions of the SALT is the time-averaged modal output power $\mathcal{P}_\mu$. This can be obtained in two ways. First it can be calculated from the asymptotic out-going fields, which are directly calculated in some numerical approaches \cite{TSG Science}. Alternatively, the Poynting flux through a loop enclosing a 2D cavity can be converted into an area integral, which gives the convenient expression:
\begin{equation}
  \mathcal{P}_\mu = \frac{k_\mu}{2\pi} \, \int_C d^2r\,
\left\{\frac{\Gamma_\mu D_0 F(\vec{r})}{1 + h(\vec{r})} -
\textrm{Im}[\epsilon(\vec{r})]\right\} \, |\Psi_\mu(\vec{r})|^2.
  \label{outputpower}
\end{equation}
A more detailed discussion and derivation of the modal output power is given in Appendix \ref{sec:outputpower}.

\subsection{Threshold Lasing Modes and Constant-Flux States}
\label{Threshold Lasing Modes and Constant-Flux States}

The lasing equation (\ref{TSG1a}) always admits the trivial solution $\Psi =
0$.  Below the first lasing threshold, this is the only self-consistent
solution.  As $D_0$ is gradually increased from zero, at some value there
emerges an additional self-consistent solution, consisting of a single lasing
mode $\Psi_\mu^{(t)}(\vec{r})$.  Right at threshold, this mode has
infinitesimal amplitude, $\Psi_\mu^{(t)}(\vec{r}) \rightarrow 0$.  Hence, the
hole-burning term $h(\vec{r})$ is negligible and (\ref{TSG1a}) reduces to a
linear equation:
\begin{equation}
  \left[\nabla^2 + \left(\epsilon_c(\vec{r}) + \gamma_\mu D_0\,F(\vec{r}) \right)k_\mu^2\right] \Psi_\mu^{(t)}(\vec{r}) = 0. \label{threshold differential equation}
\end{equation}
Note that the second term in parentheses, which we will refer to as $\epsilon_g(\vec{r})$, is simply the linear amplifying dielectric function of the pumped gain medium.  As shown in the following sections, this equation has a discrete set of non-trivial solutions, specified by the two positive real numbers, $(D_0^\mu, k_\mu^{(t)})$, the threshold values of the pump and lasing frequency.  Each of these solutions would be a perfectly valid lasing mode at threshold for that specific pump value, assuming that all other modes are suppressed for some reason.  We refer to this set of functions with their corresponding frequencies as the Threshold Lasing Modes (TLMs).  They can be thought of as the non-interacting lasing modes, i.e. the modes that would turn on in the absence of modal interactions, and their thresholds $D_0^\mu$ are the non-interacting thresholds.

\begin{figure}
  \centering\includegraphics[width=7cm]{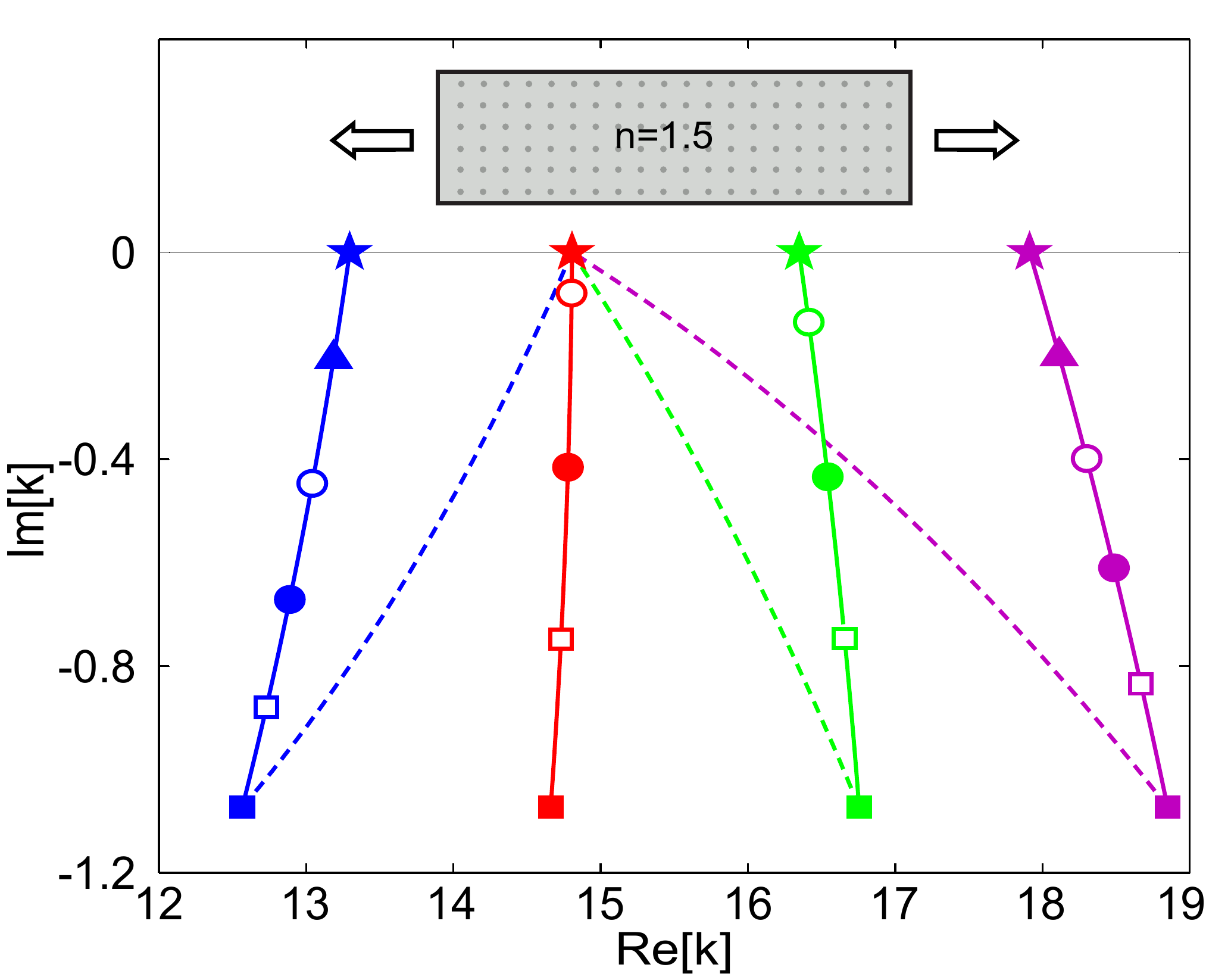}
  \caption{(Color online) Trajectories of scattering matrix poles with increasing pump strength $D_0$. Inset: Schematic of the 1D slab resonator used. Its length $L=1$ and index $n=1.5$. Grey dots indicate that the pump covers the whole resonator. Solid curves show the pole trajectories when the gain-induced dielectric constant $\epsilon_g$ is given by Eq.~(\ref{threshold differential equation}), with gain parameters $k_a=15/L$ and $\gper=3/L$.  Different symbols lying along each trajectory represent different pump strengths: $D_0=0$ (filled squares), $0.1$ (open squares), $0.2$ (filled circles), $0.3$ (open circles), and $0.4$ (filled triangles).  At $D_0=0$, the poles are the resonances of the passive cavity, which all have the same imaginary part in this case. Stars indicate the real frequencies $k_\mu^{(t)}$ of the corresponding TLMs, which arise for different pump values, $D_0=D_0^\mu$ in general.  In Eq.~(\ref{TLM basis}), we associate a basis set of TCF states with each TLM; each TCF corresponds to adding a different gain dielectric function to the medium, which pulls a different pole through the same $k_\mu^{(t)}$. The dashed lines show this process for these four poles; here we define an increasing dielectric constant $\epsilon_g$ by  $\epsilon_g = s\eta_m, (0\leq s\leq1)$, where $\eta_m$ is the TCF eigenvalue introduced in Eq.~(\ref{TLM basis}) at the frequency of the first lasing mode. The dashed and solid lines for the first mode (red) coincide.} \label{fig:smatPoles}
 \end{figure}

There is another interesting interpretation of the TLMs.  The linear wave equation (\ref{threshold differential equation}) defines an electromagnetic scattering matrix which gives the out-going wave amplitudes in terms of the incident wave amplitudes.  The outgoing-only boundary condition implies that the relevant solutions correspond to poles of this S-matrix, i.e.~eigenvectors with eigenvalue tending to infinity.  When $D_0 = 0$, these poles are just the resonances of the passive cavity defined by the wave equation
\begin{equation}
  \left[\nabla^2 + \epsilon_c(\vec{r}) k^2 \right] \psi(\vec{r}) = 0 \label{S-matrix pole}
\end{equation}
with an out-going boundary condition. If the cavity is lossless and $D_0=0$, then the corresponding S-matrix is unitary; otherwise it is not flux-conserving. For {\it any} cavity in equilibrium (i.e. lossless or absorbing) these solutions only exist for complex $k$, with $\textrm{Im}(k) < 0$; hence, outside $C$, these modes grow exponentially towards infinity, which means that they are not physical realizable \cite{TS}.  When $D_0 >0$, the dielectric function in (\ref{threshold differential equation}) is not merely the passive $\epsilon_c(\vec{r})$, but includes a complex non-equilibrium amplifying contribution
$\epsilon_g(\vec{r})$ from the gain medium, whose effect is to move the poles ``upwards'' towards the real axis (see Fig.~\ref{fig:smatPoles}).  The non-interacting thresholds associated with the TLMs are the values of the pump that move the pole corresponding to each resonance onto the real axis, making it a physically possible threshold lasing mode.

In the real system, once the pump reaches the smallest of these thresholds the solution with $D_0 = D_0^{\textrm{min}}$ turns on.  This mode then begins to contribute to the hole-burning term in (\ref{TSG1a}).  For all higher pump values this term induces nonlinear interactions by reducing the inversion, raising the thresholds for the higher modes, and in general changing both their frequencies and spatial distributions.  Thus, above the first lasing threshold we face a set of coupled, nonlinear differential equations (\ref{TSG1a}), for the unknown {\it interacting} lasing modes $\Psi_\mu(\vec{r})$ and frequencies $k_\mu$.  From a practical standpoint, the most efficient way to solve this problem is to characterize these modes with a tractably small set of variables, by expanding them in an appropriate choice of basis functions.  The original formulation of the SALT employed the following basis set:
\begin{align}
  \begin{aligned}
    \left[\nabla^2 + \epsilon_c(\vec{r}) K_{n}^2(k) \right]
    \varphi_{n}(\vec{r},k) &= 0, \quad \vec{r} \in C\\ \left[\nabla^2 +
    n_0^2k^2\right] \varphi_{n}(\vec{r},k) &= 0, \quad \vec{r} \notin C
    \label{CF modes}
  \end{aligned}
\end{align}
where $K_{n}$ are complex and $k$-dependent. The basis states $\varphi_{n}(\vec{r},k)$ were called the ``constant flux'' (CF) states, and they satisfy an out-going boundary condition at the cavity boundary $\partial C$ \cite{bibnote1}.  Within $C$, they obey a wave equation with the complex frequency $K_n$, analogous to (\ref{S-matrix pole}).  Outside, they obey a wave equation with real frequency $k$, and are required to be outgoing at infinity. The total electromagnetic energy flux outside $C$ is conserved, as it must be for a physical mode.

The ``constant-flux'' condition outside $C$ can be satisfied by a variety of complete non-Hermitian basis sets.  The specific CF basis (\ref{CF modes}), used in Refs.~\cite{TS,TSG PRA, TSG Science,TSG Nonlinearity}, was chosen because of its similarity to the equation defining the resonances; it differs from (\ref{S-matrix pole}) only by having real $k$ outside the cavity.  If the cavity dielectric $\epsilon_c$ is constant and the pump is uniform ($F=1$), then each TLM is a CF state; and (ii) the complex frequency $K_n (k)$ of the CF state is very close to the complex frequency of a passive cavity resonance \cite{TS,TSG PRA}.  To be precise, the CF frequency corresponding to a TLM is
\begin{equation}
  K_{n=\mu}(k_\mu^{(t)}) = \left[1 + \frac{\gper D_0 / \epsilon_c}{k_\mu^{(t)}- k_a + i\gamma_\perp}\right]^{-\frac{1}{2}}\,k_\mu^{(t)}.
\end{equation}
If we define $K_{n} = q_n - i \kappa_n$ (suppressing k-dependence) and assume that the the lasing frequency is close to the atomic frequency, it is easily shown \cite{TSG PRA} that
\begin{equation}
k_\mu^{(t)} = k_a + \frac{\gamma_\perp q_\mu}{\gamma_\perp + \kappa_\mu},
\end{equation}
which is the familiar line-pulling formula for the single mode lasing frequency \cite{Haken}, with $q_\mu,\kappa_\mu$ playing the role of the cavity frequency and linewidth.  This emphasizes the relationship of the SALT to earlier theories that identified lasing modes with passive cavity resonances.

When the cavity dielectric function $\epsilon_c$ and/or the pumping profile $F$ is non-uniform, the TLMs are not given by a single CF state, and each must be written as a superposition of CF states.  In Ref.~\cite{TSG Science}, it was found that practical SALT calculations can be performed using a basis of $20-50$ CF states.  However, when the pumping is non-uniform, the rate of convergence of the CF basis set is poorer.  Although the CF state definition (\ref{CF modes}) takes $\epsilon_c(\vec{r})$ into account, it does not include the pump profile $F(\vec{r})$ as an independent parameter; effectively, these CF states correspond to a pump profile proportional to $\epsilon_c(\vec{r})$.

The above drawbacks motivate us to introduce a new basis set for the SALT equations.  These basis functions are still CF states, in the sense that they obey the real-$k$ out-going boundary conditions.  However, their definition accounts for non-uniformity in both the cavity dielectric function and the pump profile, allowing us to assign a basis set to each TLM, with {\it one} of the basis functions exactly equal to the TLM.  We will see that the nonlinear above-threshold solutions can be expanded with a minimum number of these basis functions, resulting in a marked improvement in the performance of the SALT.

To avoid confusion, we henceforth refer to the original CF states (\ref{CF modes}) as Uniform Constant Flux (UCF) states, and the new basis states as Threshold Constant Flux (TCF) states.

\section{Threshold Constant Flux states and SALT equations in CF bases}
\label{Basis functions section}

\subsection{Threshold Constant Flux States}

We define the TCF states by:
\begin{align}
  \begin{aligned}
    \left[\nabla^2 + \Big(\epsilon_c(\vec{r}) + \eta_{n}(k) \,
      F(\vec{r})\Big) k^2 \right] u_{n}(\vec{r},k) &= 0, \quad
    \vec{r} \in C\\ \left[\nabla^2 + n_0^2k^2\right] u_{n}(\vec{r},k)
    &= 0, \quad \vec{r} \notin C
    \label{TLM basis}
  \end{aligned}
\end{align}
where $\eta_{n}$ are complex and $k$-dependent, and $u_{n}(\vec{r},k)$ are outgoing with frequency $k$ at infinity. $F(\vec{r})$ is the spatial pump profile defined in (\ref{eq:F(r)}). For each $k$, there exists a discrete set $\{u_{n}(\vec{r},k), \eta_{n}(k)\, |\, n = 1, 2, \cdots\}$ of solutions to (\ref{TLM basis}).  We refer to $\eta_{n}$ as the TCF eigenvalue, for reasons that will become clear.

Like the UCF frequencies $K_n$, the TCF eigenvalues $\eta_{n}(k)$ are complex, not real, due to the open (non-Hermitian) boundary condition.  One can show that $\textrm{Im}[\eta_{n}(k)] < 0$, which implies amplifying behavior similar to the condition ${\rm Im}[K_n] < 0$ for the UCF states.  In (\ref{TLM basis}), $\eta_{n}(k)F(\vec{r})$ plays the role of a complex amplifying dielectric function with the same spatial profile as the pump, so $\eta_{n}(k)$ physically is the scale of the amplifying dielectric constant necessary for that TCF to
reach threshold and emit at wavevector $k$. As previously stated, if we choose $k=k_\mu^{(t)}$, then one of the basis functions matches the solution $\Psi_\mu^{(t)}$ for the threshold lasing equation (\ref{threshold differential equation}):
\begin{equation}
  u_{n}(\vec{r},k_\mu^{(t)}) = \Psi_\mu^{(t)}(\vec{r}),
  \label{TLM}
\end{equation}
for index $n$ such that
\begin{equation}
  \eta_{n}(k_\mu^{(t)}) = \frac{\gper D_0^\mu}{k_\mu^{(t)} - k_a + i\gamma_\perp}. \label{threshold_condition}
\end{equation}
Note that TLMs and TCF states both satisfy linear equations, and hence have no overall scale, so the same normalization must be assumed in (\ref{TLM}).  Thus each infinite TCF basis set is associated with one true TLM, indexed by $\mu$.  Slightly above threshold, this one TCF state serves as a very good approximation for the first lasing mode.  Well above threshold, the lasing mode must be constructed from a superposition that includes the other TCF states (the lasing frequency $k_\mu$ will also change slightly from its threshold value as the pump increases, and the TCF states will adjust accordingly).  As noted, these other TCF states correspond to different values of $\epsilon_g$ that would also lead to lasing at $k_\mu$, values that are not realized by the two-level gain medium of the MB equations.  In the S-matrix picture, they correspond to moving a {\it different} pole through the real axis at $k_\mu^{(t)}$, as indicated in Fig.~\ref{fig:smatPoles}.  Higher lasing modes can likewise be expanded using TCF states with different $k_\mu$.

The TCF states are not power-orthogonal, but obey a self-orthogonality relation:
\begin{equation}
  \int_C d^d\!r \; F(\vec{r})\; u_{n}(\vec{r},k)\, u_{n'}(\vec{r},k) =
  \delta_{nn'}.
  \label{biorthogonal}
\end{equation}
We use the superscript $d$ to indicate the dimensions of the system here and in the following discussion. We assume degenerate $\eta_n$'s are handled, as usual, by choosing the basis so that (\ref{biorthogonal}) is satisfied.  It follows that any sufficiently regular function having the same out-going boundary condition (with frequency $k$) can be expanded in the TCF basis $\{u_{n}(\vec{r},k)\}$. For the uniform case, the UCF and TCF states are the same, with eigenvalues related by
\begin{equation}
\eta_n (k) = \epsilon_c (K_n^2/k^2-1). 
\end{equation}

Interestingly, basis states of the UCF type were first defined and used by Kapur and Peierls \cite{Kapur} in the context of nuclear decay, long before their introduction to optical physics by T\"ureci \textit{et~al.} \cite{TS}.  The $k$-dependence of the Kapur-Peierls basis set was considered inconvenient, and it was largely superseded by the use of S-matrix resonances, which do not form a complete basis set but are useful when single-pole approximations are valid (and the amplifying behavior at infinity is ignored) \cite{wigner}.  In our present situation, the appearance of internal amplifying eigenvalues is much more natural, for there is truly a gain medium within the cavity!  The Kapur-Peierls (CF) approach, and not the resonance approach, is thus the natural one for describing the laser; and with the availability of modern computers, the fact that the basis is $k$-dependent does not pose any serious difficulty.

\subsection{Threshold Lasing Conditions}
\label{threshold discussion}

\begin{figure}
  \centering\includegraphics[width=7cm]{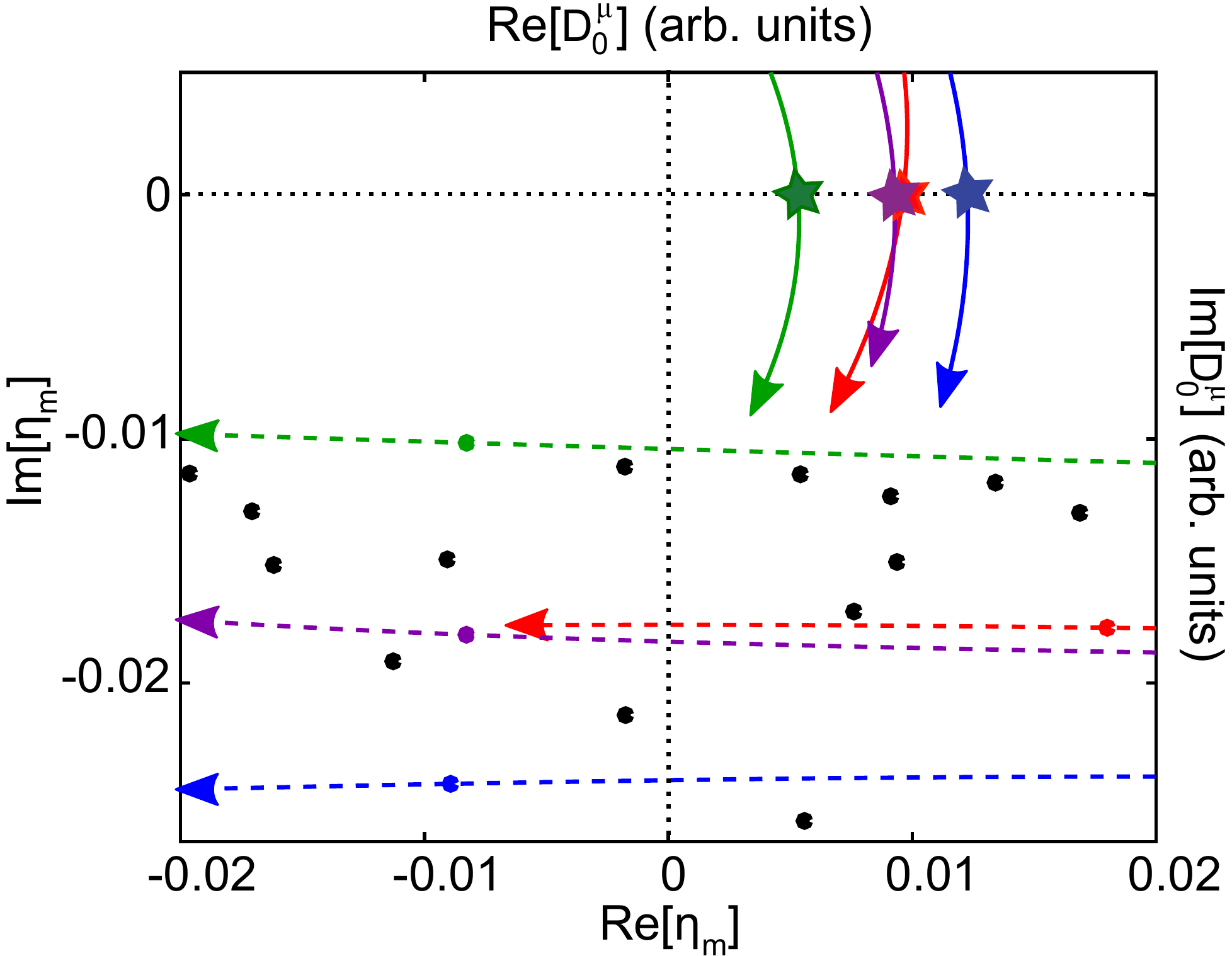}
  \caption{(Color online) Relation between TCF eigenvalues $\eta_n$ and the threshold inversion $D_0^\mu$. Dots show the complex TCF spectrum $\eta_n(k)$ for a random laser. Dashed curves show the $k$-dependence of four of these TCF  eigenvalues, with arrows indicating increasing $k$.  Solid curves show the corresponding $D_0^\mu(k)$ obtained from Eq.~(\ref{threshold_condition}), with $\gamma_\perp = k/60$.  A TLM occurs when one $D^\mu_0(k)$ hits the real axis, as indicated here by stars. The green colored TCF eigenvalue has a small imaginary part and a real part fairly close to zero; thus it leads to the lowest threshold (smallest $D_0^\mu$), consistent with the discussion in the text.} \label{tlmfreq plot}
\end{figure}

We have seen that the first TLM, having frequency $k = k_\mu^{(t)}$,
corresponds exactly to a single TCF state $u_{n}(\vec{r},k_\mu^{(t)})$, and that the other TCF states  must be included above threshold, even though they are not possible TLMs for the actual system.  We can find the first TLM by computing the TCF states and $\{\eta_n(k)\}$ over a range of frequencies close to the gain center $k_a$.  For  for a fixed choice of
$(\eta_n,k)$, Eq.~(\ref{threshold_condition}) will yield a complex (unphysical) value for $D_0$, but when $D_0$ passes through the real axis at $k = k_\mu^{(t)}$, the value of $\eta_{n=\mu}$ defines a TLM according to (\ref{TLM}--\ref{threshold_condition}) (see Fig. \ref{tlmfreq plot}). The first lasing mode is then the TLM with the smallest $D_0^\mu$.  The other TCFs for that TLM are $\{u_{m}(\vec{r},k_\mu^{(t)})\,|\,m \ne n\}$.

To identify which $\eta_n$ will generate low threshold TLMs, for real $D_0$,
we can rewrite (\ref{threshold_condition}) explicitly as
\begin{eqnarray}
k &=& k_a - \frac{\textrm{Re}[\eta_{n}(k)]}{\textrm{Im}[\eta_{n}(k)]}\,  \gamma_\perp,  \label{threshold cond 1} \\
D_0 &=& -\textrm{Im}[\eta_{n}(k)] \, \gamma_\perp\, \left[ 1+ \left(\frac{k-k_a}{\gamma_\perp}\right)^2\right], \label{threshold cond 2}
\end{eqnarray}
with $k = k_\mu^{(t)}$.  From the expression in brackets in (\ref{threshold cond 2}), $|k-k_a|$ should be as small as possible --- and hence, via (\ref{threshold cond 1}), so should $|\textrm{Re}(\eta_n)|$.  From the prefactor in (\ref{threshold cond 2}), $|\textrm{Im}(\eta_n)|$ should also be small, and this condition becomes relatively more important than the first when $\gamma_\perp$ is large, i.e.~the gain curve is broad.  Thus, the relevant TCF states are those lying within a ``window'' around $\textrm{Re}(\eta) \approx 0$ of width $\sim \gamma_\perp$; within this window, states with $\textrm{Im}(\eta)$ closest to zero (i.e. requiring the least gain) are favored.  This analysis agrees with the numerical results shown in Fig.~\ref{tlmfreq plot}.

We can also express the threshold lasing mode in terms of the UCF modes
(\ref{CF modes}).  As noted in Sec.~\ref{Threshold Lasing Modes and Constant-Flux States}, for non-uniform $\epsilon_c$ and/or $F$ it is necessary to use a superposition of UCF modes:
\begin{align}
\begin{aligned}
  &\Psi_\mu(\vec{r}) = \sum_n \alpha_n^\mu \, \varphi_n(\vec{r}), \\
  &\alpha_n^\mu = \frac{\gamma_\mu D_0 k^2}{K_n^2-k^2} \sum_{n'} \int_Cd^dr\,
    F(\vec{r})\, \varphi_n(\vec{r})\, \varphi_{n'}(\vec{r})  \alpha_{n'}^\mu,
  \label{CF threshold equation}
\end{aligned}
\end{align}
with $k = k_\mu^{(t)}$.  This formulation of the SALT was used in Ref.~\cite{TSG Science} to analyze 2D random lasers.

\begin{figure}
\centering
\includegraphics[width=7cm]{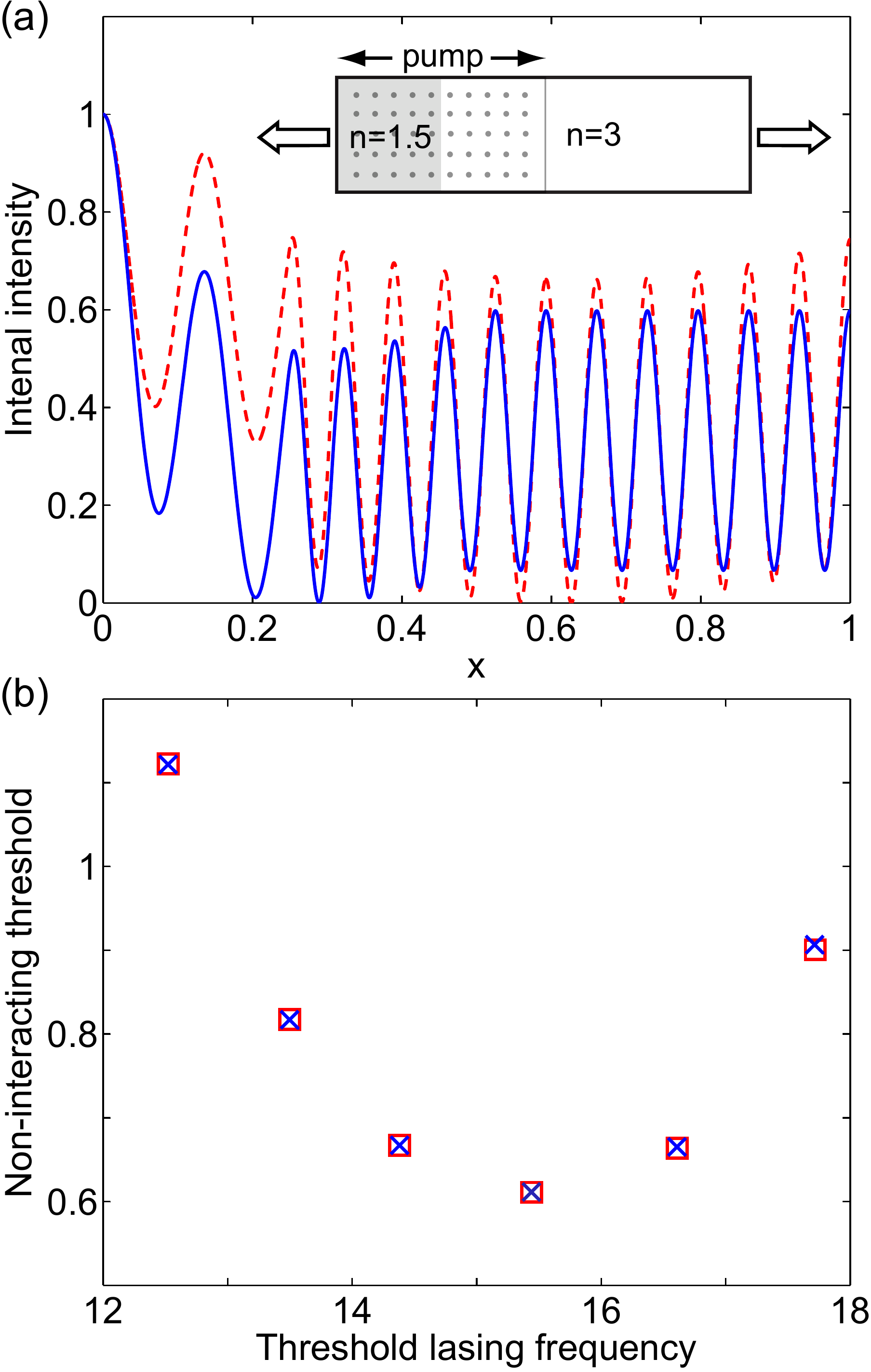}
  \caption{(Color online) (a) Spatial profile of the first threshold lasing mode of a 1D slab resonator of of length $L=1$ (solid blue curve).  This TLM corresponds exactly to a TCF state.  The matching UCF state, having the largest overlap with this lasing mode, is shown for comparison (red dashed curve). Inset: Schematic of the resonator. The refractive index is $n=1.5$ for $0<x<0.25$, and $n=3$ for $0.25<x<1$. The gain center is $k_a=15/L$, and only the left half of the cavity is pumped (indicated by the grey dots).  Both states are normalized to unity at $x = 0$.    (b) Lasing frequencies $k_\mu^{(t)}$ and non-interacting threshold values $D_0^\mu$ of the six TLMs with the lowest thresholds. Crosses are given the TCF solutions, and squares are the UCF solutions with 20 UCF states. } \label{fig:CFvsMLG1d}
\end{figure}

To illustrate the advantage of the TCF basis for non-uniform cavity dielectric function $\epsilon_c$ and pumping profile $F$, we study a 1D resonator of length $L=1$. The refractive index is $n=1.5$ for $0<x<0.25$, and $n = 3$ for $0.25<x<1$.  Only the left half of the cavity ($0<x<0.5$) is pumped. The TCF state corresponding to the first TLM, with threshold $D_0=0.611$, is plotted in Fig.~\ref{fig:CFvsMLG1d}(a), along with the UCF state making the largest contribution to this TLM.  The TCF state is tailored to the pump profile and is only amplified in the pumped region, whereas, as already noted, the UCF states have no knowledge of the pump profile and exhibit amplification within the entire cavity, including the unpumped region.  The UCF state shown in the figure represents only $54.0\%$ of the total weight in this superposition \cite{weightMethod}.

In order to reproduce the actual TLM, we must superpose many UCF states to cancel the amplification in the unpumped region.  In Fig.~\ref{fig:CFvsMLG1d}(b) we plot the lasing frequencies $k_\mu^{(t)}$ and (non-interacting) thresholds $D_0^{\mu}$ of the six TLMs with the lowest thresholds obtained by solving (\ref{threshold_condition}) and (\ref{CF threshold equation}) with 20 UCF states, respectively.  The largest deviation between the TCF and UCF thresholds is $0.68\%$, and the frequency differences are below $0.1\%$.

\begin{figure}
\centering
\includegraphics[width=7cm]{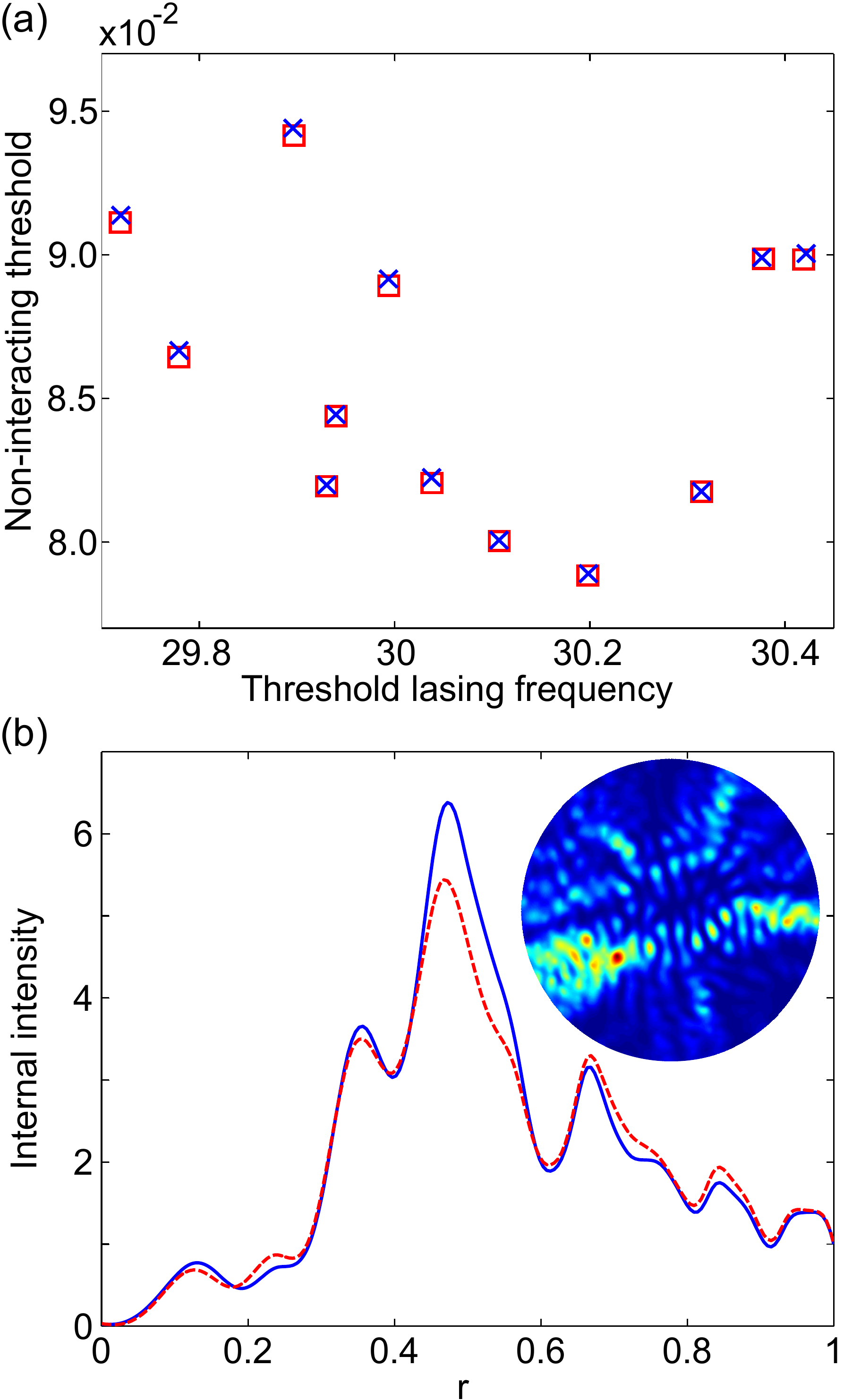}
\caption{(Color online) Comparison of TCF and UCF results for 2D random lasers.  (a) Lasing frequencies $k_\mu^{(t)}$ and non-interacting threshold values $D_0^\mu$ of the twelve lowest TLMs, for a white-noise pump covering the whole cavity.  The pump profile is given by Eq.~(\ref{white noise source}).  Crosses show the TCF solutions for (\ref{threshold cond 1}) and (\ref{threshold cond 2}), and squares show the UCF solutions for (\ref{CF threshold equation}) with 50 UCF states.  (b) Spatial intensity profiles for the first threshold lasing mode of a partially pumped random laser, with $F = 1$ for $r < R/2$ and $F = 0$ for $r > R/2$.  The intensity is plotted along the line $\theta = 225^\circ$.  The solid curve shows the TCF solution, and the dashed curve shows the UCF solution computed from 200 UCF states.  Inset: false-color intensity plot from the superposition of 200 UCF states; the TCF intensity plot, which is not shown, looks similar.  } \label{fig:RL_th}
\end{figure}

In more complex lasers, e.g. the 2D random lasers of Ref.~\cite{TSG Science}, a still larger UCF basis set is required to achieve results comparable with the TCF basis.  In Fig.~\ref{fig:RL_th}, the cavity $C$ is defined by a disk of radius $R=1$, in which we randomly place 600 dielectric particles of radius $\sim R/80$ and index $n=1.2$.  In Fig.~\ref{fig:RL_th}(a), we subject the entire cavity to a white noise pump,
\begin{equation}
  F(\vec{r}) = 1 + \xi(\vec{r}),
  \label{white noise source}
\end{equation}
with max$|\xi(\vec{r})|=0.3$.  We find that 50 UCF states must be included in the UCF expansion in order to achieve good agreement between the threshold solutions of (\ref{CF threshold equation}) and the TCF predictions (\ref{threshold cond 1}) and (\ref{threshold cond 2}).  When we pump only part of $C$ (keeping the scatterer configuration fixed), more UCF states are needed to correctly reproduce the TLMs, even in the absence of the pump noise.  In Fig.~\ref{fig:RL_th}(b), the pump covers a central area of radius $R/2$.  We find that a superposition of 200 UCF states is required to generate a TLM whose false-color intensity plot (inset) is indistinguishable by eye from the corresponding TCF state (not shown).  Even with this many UCF states, the computed TLM intensity profile still differs significantly from the exact (TCF) profile when plotted along any arbitrary direction, as shown in the main figure.  The reason so many UCF states are required is that the TLM (and the corresponding TCF state) is amplified only up to the boundary of the pump region; whereas each UCF state, like the uniform pumped system, is amplified up to the boundary of $C$ \cite{TSG Science}.  Using these 200 UCF states, the calculated mode threshold and frequency are $D_0 = 0.142$ and $k=30.011$; the exact TCF results, from (\ref{threshold cond 1}) and (\ref{threshold cond 2}), are $D_0 = 0.140$ and $k=30.006$.

\subsection{Above-threshold Lasing Modes}
\label{Above-threshold lasing modes}

Above threshold, each lasing mode can be efficiently expanded as a superposition of TCF states:
\begin{equation}
  \Psi_\mu (\vec{r}) = \sum_n a_n^\mu \, u_{n}(\vec{r},k_\mu).
  \label{eq:TCFexpansion}
\end{equation}
This expansion automatically satisfies the appropriate out-going free wave equation outside $C$. By inserting the above expansion into (\ref{TSG1a}), we write the latter as
\begin{equation}
\frac{D_0\Psi_\mu(\vec{r})}{1 + h(\vec{r})} = \sum_n \frac{\eta_{n}}{\gamma_\mu} a^\mu_n \,u_{n}(\vec{r}). \label{eq:TSG_TLM0}
\end{equation}
Following the procedures used in Ref.~\cite{LiThesis}, we multiply both sides of (\ref{eq:TSG_TLM0}) by $F(\vec{r})u_{n'}(\vec{r})$, integrate $\vec{r}$ over $C$, and invoke the self-orthogonality property (\ref{biorthogonal}), to find the SALT equation in the TCF basis:
\begin{align}
\begin{aligned}
  & D_0 \sum_{n'} \mathcal{T}_{nn'} a_{n'}^\mu = \,a_n^\mu, \\
  &\mathcal{T}_{nn'} \equiv \frac{\gamma_\mu}{\eta_{n}}\, \int_Cd^dr\;
    \frac{F(\vec{r})\, u_n(\vec{r})\, u_{n'}(\vec{r})}{1+h(\vec{r})}.
  \end{aligned}\label{eq:SALT_TCF}
\end{align}
Eq.~(\ref{eq:SALT_TCF}) is a set of nonlinear fixed-point equations above threshold, one for each lasing mode.  In general, the complex matrix $\mathcal{T}_{nn'}(k)$, which we refer to as the lasing map, has complex eigenvalues.  Because the pump strength $D_0$ is a real variable, the unknown lasing frequency $k_\mu$ must be such that one of its (nonlinear) eigenvalues is real and equal to $1/D_0$.  This is achieved by tuning $k_\mu$ to find the values at which the different eigenvalues corresponding to the different modes cross the real axis, as follows.  (This procedure is the same as for the UCF basis, and was described in Ref.~\cite{TSG Nonlinearity}).  The first threshold and lasing frequency are found simply by solving (\ref{threshold cond 1}) and (\ref{threshold cond 2}) self-consistently, as described in the previous section; these equations are the diagonal form of (\ref{eq:SALT_TCF}) at threshold.  The associated TLM is proportional to this solution, with vanishing overall amplitude.  We then increase $D_0$ in small increments, and use the solution for the smaller pump value as a starting point for the nonlinear solver.  At each step, the nonlinear solver adjusts the frequency $k_\mu$ so that the corresponding eigenvalue of $\mathcal{T}_{nn'}(k_\mu)$ is real.  From the modified lasing map, which includes the hole-burning term, we can determine if a second mode has reached its (interacting) threshold \cite{TSG Nonlinearity}.  A similar procedure works for third and higher modes, and has been shown to work for systems as complex as a 2D random laser with eight modes turned on \cite{TSG Science}.

In earlier works, the lasing map was written in the UCF basis.  This has the same form as (\ref{eq:SALT_TCF}), with a slightly different matrix operator:
\begin{align}
\begin{aligned}
  T_{nn'} \equiv  \frac{\gamma_\mu \, k_\mu^2}{K_{n}^2-k_\mu^2}\,
  \int_Cd^dr\,\frac{F(\vec{r})\, \varphi_n(\vec{r})\, \varphi_{n'}(\vec{r})}{1+h({\vec{r})}} .
  \label{eq:SALT_UCF}
\end{aligned}
\end{align}
At threshold ($h \rightarrow 0$), we recover the threshold lasing equation (\ref{CF threshold equation}).  The solution algorithm is identical to that for the TCF map, except that the full matrix solution must be performed even at the first threshold since the UCF map is not diagonal.

\begin{figure}
\centering
\includegraphics[width=7cm]{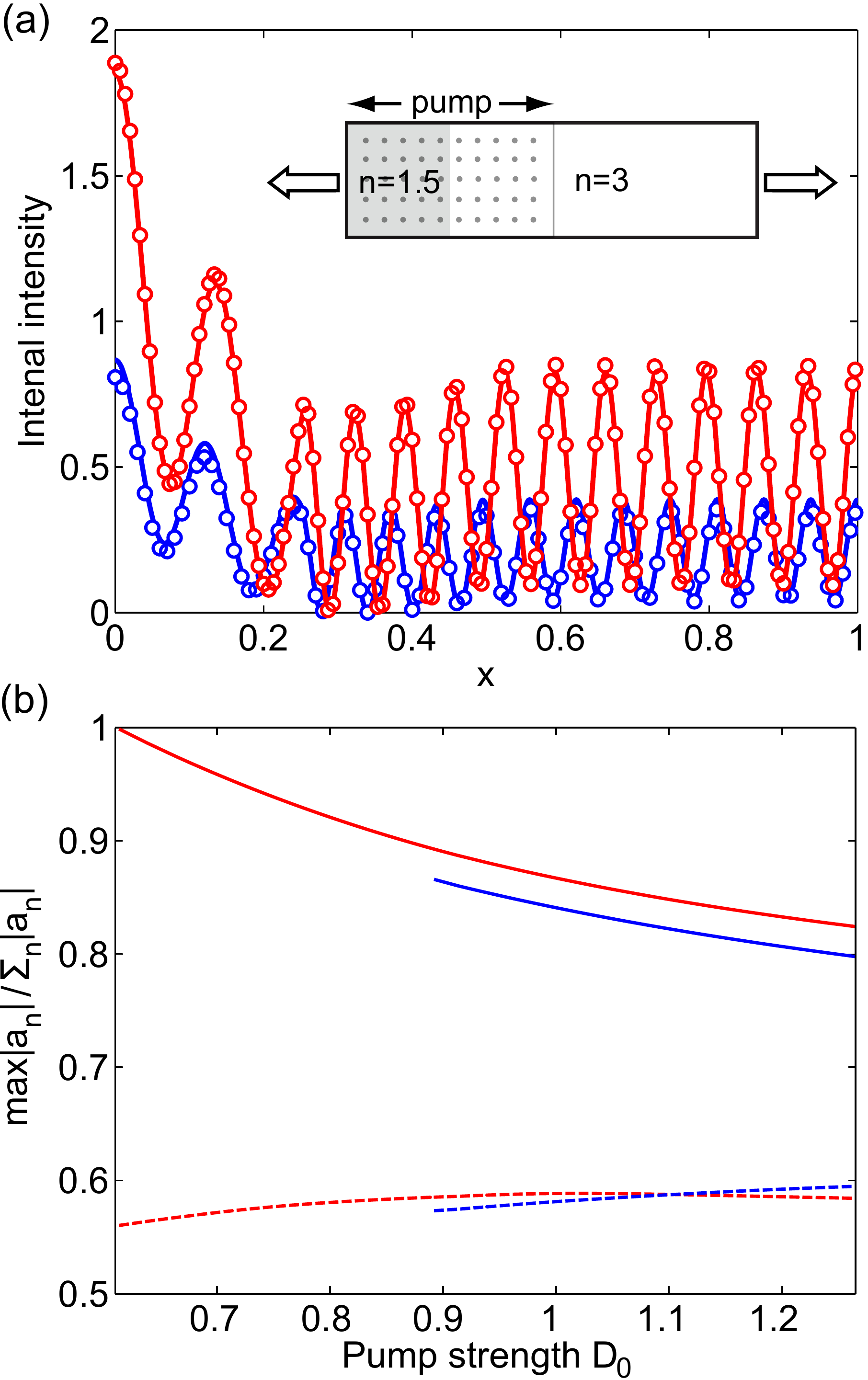}
  \caption{(Color online) (a) Spatial profiles of two lasing modes above threshold in a 1D slab resonator. Inset: Schematic of the resonator. The properties of the resonator are given in the caption of Fig.~\ref{fig:CFvsMLG1d}. The pump strength $D_0=1.264$ is slightly higher than twice the first threshold. The solid lines and circles are the results of (\ref{eq:SALT_TCF}) and (\ref{eq:SALT_UCF}), respectively, both using 20 basis functions. (b) Weights of the largest expansion coefficients of both modes in the TCF (solid curves) and UCF (dashed curves) bases. The second mode has an interacting threshold
$D_0 = 0.89$.} \label{fig:aboveTH}
\end{figure}

Fig.~\ref{fig:aboveTH} compares the lasing modes obtained from (\ref{eq:SALT_TCF}) and (\ref{eq:SALT_UCF}) for the 1D slab resonator that we studied earlier in Fig.~\ref{fig:CFvsMLG1d}.  For $D_0=1.264$, there are two lasing modes.  (This value of $D_0$ is approximately twice the first lasing threshold, $D_0^{\mu=1}=0.611$.)  Using 20 basis functions for both methods, we find good agreement in the predicted spatial profiles.  Fig.~\ref{fig:aboveTH}(b) shows the largest expansion coefficients of the two modes in the TCF and UCF bases.  We find that both modes retain a dominant component in the TCF basis, even when the system is significantly above threshold.  As the pump strength increases, the spatial hole burning term changes $\epsilon_g(\vec{r})$, so the weights of the dominant components in the TCF basis gradually decrease, but they remain larger than $80\%$ in the calculated range.  In contrast, the largest components of the two modes in the UCF basis are less than $60\%$.

We remark that we could in principle absorb the hole-burning denominator $1/[1+h(r)]$, calculated at the pump strength $(D_0 - \delta D_0)$, into the profile function $F(\vec{r})$, to produce an even better set of modified TCF states for the nonlinear calculation at $D_0$. This is essentially an alternative means of solving the non-linear problem by keeping the self-consistent equation almost diagonal in an evolving basis; however, it is usually too computationally expensive to recompute the TCF states this way.

\section{Approximate analytic solution of the SALT equations}
\label{Approximate analytic solution of SALT equations}

\subsection{Alternative Fixed-point Equation}

Analyses of the MB equations, either in the single-mode or multi-mode lasing regime, almost always employ the near threshold approximation, in which the infinite-order nonlinearity of Eqs.~(\ref{TSG1}-\ref{TSG2},\ref{eq:SALT_TCF}) is truncated at cubic order to give a near-threshold approximation to the solution. (An exception to this is the work of
Mandel and coworkers \cite{Mandel1,Mandel2} discussed in Appendix B).
 Based on this cubic approximation, and the approximation of a closed cavity, Haken and Sauermann (HS) long ago derived a set of constrained linear equations for the modal intensities in the multimode regime \cite{HS}.  The HS equations have been studied further \cite{Haken,HakenFu}, and have been used to analyze random and complex lasers in recent years \cite{Misirpashaev,TS05}.  However, the results are unsatisfactory, as shown by T\"ureci {\it et al}. \cite{TS}.  The cubic nonlinearity in the HS equations leads to a saturation of modal intensities, in disagreement with the linear increase expected on general grounds, and found by more exact treatments \cite{TSG PRA,TSG OE}.  It also allows many more modes to turn on than in the more exact treatments \cite{TS05,TSG OE}.  These failures are unsurprising, as the HS theory is being applied to a regime well above the first threshold, where the cubic approximation is poor.  Generalizing the equations from cubic to higher orders rapidly becomes unmanageable, since higher powers of the intensity generate many more interaction coefficients to take into account.

In Ref.~\cite{TS}, it was shown that in the limit of large hole-burning, $h(\vec{r}) \gg 1$, the SALT correctly predicts mode intensities growing linearly with $D_0$, within the single-pole approximation to be discussed below.  In this current section, we will derive an alternative lasing map which, in the same approximation, allows a more complete analytic solution that demonstrates linear behavior for \textit{all} values of $h(\vec{r})$ and for multi-mode lasing.  This approximation also provides a quantitative solution for the modal intensities, slopes and interacting thresholds, in good agreement with the exact solutions of the SALT equations.  To our knowledge, these are the first results of this type, valid for arbitrarily complex cavities, to appear in the literature.

In order to develop the desired approximation, we first re-express the lasing equations in terms of the inverse of the map $\mathcal{T}_{nn'}(k)$ defined in (\ref{eq:SALT_TCF}).  This inverse map has the same fixed points, but is much more convenient to work with.  We multiply both sides of (\ref{eq:TSG_TLM0}) by $[1+h(\vec{r})]$, and repeat the steps leading to (\ref{eq:SALT_TCF}), i.e.~projecting the two sides onto the TCF basis and using the self-orthogonality property (\ref{biorthogonal}). The result is
\begin{eqnarray}
  \sum_{n'} \tau_{nn'}a_{n'}^\mu &=& D_0 \, a_{n}^\mu, \label{eq:inverse} \\
  \tau_{nn'} &=& \frac{\eta_{n'}}{\gamma_\mu}\,\left[\delta_{nn'} + h_{nn'}\right],
\end{eqnarray}
where $h_{nn'}(k) = \int_C d^d\!r\, F(\vec{r},k) \,h(\vec{r})\,u_{n}(\vec{r},k)\,u_{n'}(\vec{r},k)$.  Note that (\ref{eq:inverse}) has the same form as (\ref{eq:SALT_TCF}), but with the quantity $D_0$, which plays the role of the eigenvalue, inverted.  This implies that $\tau = \mathcal{T}^{-1}$, which can be confirmed by multiplying the two operators and using the completeness and self-orthogonality of the TCF states.

The operator $\tau$, through the term $h_{nn'}(k)$, contains only a second-order dependence on the lasing modes, in contrast to the infinite-order dependence occurring in $\mathcal{T}$.  Thus, (\ref{eq:inverse}) possesses only a cubic nonlinearity, but this is \textit{not} the same cubic nonlinearity that appears in the HS theory.  No Taylor expansion has been performed; the inverse lasing map is exact at all pump values, and we are still working with infinite-order nonlinearity in the conventional sense of using a dielectric function which contains the field to infinite order.

We could, in principle, use the inverse map $\tau$ to solve the SALT equations, in the same way that we used $\mathcal{T}$.  Preliminary investigations show that such an approach is possible, and may have some interest, but we will not pursue this further here.  Our aim is instead to introduce the ``single-pole approximation'' (SPA) into (\ref{eq:inverse}).  This gives a simple approximate solution that is very easy to implement, and yields important analytic results.

\subsection{Single-pole SALT Equations}

The single-pole approximation was introduced in  \cite{TS} to show the connection between the SALT equations, which solve the MB equations with minimal approximations (principally the RWA and the stationary inversion approximation) \cite{TSG OE}, and the HS equations which employ many more approximations.   Aside from the aforementioned cubic approximation, the HS theory assumes that the lasing mode is accurately described by a passive cavity mode.  As we have seen, even the threshold lasing mode is {\it not} a passive cavity mode: it is neither a closed cavity mode (as assumed by HS), nor a passive cavity resonance as often assumed in the literature.  Furthermore, above threshold the nonlinearity mixes in other TCF states, which changes the spatial distribution, amplitude, and frequency of the lasing mode.  This effect is quite important in low-Q cavities, such as the random lasers treated in Ref.~\cite{TSG Science}, and the full SALT theory describes this effect very well.  In high-Q cavities, the mixing in of other TCFs is much weaker, because the scattering from the gain medium is so much weaker than the scattering from the cavity itself.  Therefore, it is reasonable to assume that the lasing modes above threshold have the same spatial profile as the TLM, with an amplitude that can increase with $D_0$.  This is equivalent to taking only one term in the expansion of the cavity Green's function in the CF basis; since each term has a single pole in the complex plane, T\"ureci {\it et al}. \cite{TS} called this approach the single-pole approximation (SPA).

To be precise, the SPA assumes that
\begin{equation}
\Psi_\mu (\vec{r}) = \sum_n a_n^\mu u_n (\vec{r},k_\mu) \approx a_{n_0}^\mu u_{n_0} (\vec{r},k_\mu^{(t)}) \equiv a_\mu u_\mu (\vec{r}),
\end{equation}
where $u_{n_0}(\vec{r})$ is the TCF which is equal to the TLM at threshold and $k_\mu^{(t)}$ is the threshold value of the lasing frequency.  With this approximation the additional index $n$ is redundant and can be omitted, as we do henceforth. Thus the SPA assumes both that the lasing modes are fixed as TLMs, and that the lasing frequencies are fixed to be their threshold values.  The remaining quantities to be calculated are just the number of modes and their amplitudes $a_\mu (D_0)$ at a given pump value \cite{intensity}.  This also necessitates finding the interacting thresholds $D_{0,\rm{int}}^\mu$.

With this approximation the nonlinear matrix equation (\ref{eq:inverse}), after canceling a common factor $a_\mu$, is {\it linear} for the modal intensity $I_\mu \equiv |a_\mu|^2$:
\begin{eqnarray}
\frac{D_0}{D_0^\mu}-1 &=& \sum_\nu \Gamma_\nu\chi_{\mu\nu}I_\nu \label{eq:SPASALT} \\
\chi_{\mu\nu} &\equiv& \int d^dr\; F(\vec{r}) \; u_\mu^2(\vec{r}) \;|u_\nu(\vec{r})|^2.
\end{eqnarray}
Here $D_0^\mu = \eta_\mu/\gamma_\mu$ are the non-interacting thresholds for the TLMs, which are obtained together with $k_\mu^{(t)}$ at threshold using (\ref{threshold cond 1}) and (\ref{threshold cond 2}).  Because the frequencies of the modes are assumed to be fixed, the spectral gain factor $\Gamma_\nu$ and the ``interaction constants'' $\chi_{\mu\nu}$ are pump-independent quantities. Note also that every quantity in Eq.~(\ref{eq:SPASALT}) is real except for $\chi_{\mu\nu}$, which must have some imaginary part if the cavity is open.  This inconsistency is a consequence of the single-pole approximation; however, the higher Q the cavity, the smaller is the imaginary part, and for most cavities of interest it is acceptable to neglect this imaginary part.  Henceforth we will use the approximation $\chi_{\mu\nu} \approx
{\rm Re}[\chi_{\mu\nu}]$, and simply denote the real part with the same symbol.  With this approximation, the matrix $\chi_{\mu\nu}$ is real and has positive elements.

The above result, which we will term the SPA-SALT, bears a remarkable resemblance to the HS equations.  Those equations take the form
\begin{equation}
1-\frac{\kappa_\mu}{D_0} = \sum_\nu \Gamma_\nu \chi_{\mu\nu}I_\nu. \label{eq:HS}
\end{equation}
It can be shown that the cavity decay rate $\kappa_\mu$, a quantity inserted by hand in that theory, is simply $D^\mu_0$ in the SALT, which is calculable once the cavity and pump profile are given.  The coupling matrix in the HS equations has exactly the same form as in the SPA-SALT, except that HS used closed cavity modes (not the real part of
of the open cavity TLMs), and did not take into account the pump profile, $F(\vec{r})$.  However, the different dependence of (\ref{eq:HS}) on $D_0$ in comparison to (\ref{eq:SPASALT}) leads to very different behavior at large pump values.  For pumps near the first threshold, the two equations are approximately the same, but at large pump it is easy to show that the modal intensities in the HS theory saturate to a constant, whereas in the SPA-SALT they are proportional to $D_0$.  (It should be noted that $D_0$ in the MB equations, which we refer to as the pump, is actually the equilibrium value of the inversion in the absence of laser emission.  When one has a multi-level laser with a true pump between upper and ground levels which are distinct from the lasing transition, the quantity $D_0$ is a function of the pump which is linear at small pumps, but saturates eventually, and is bounded by the value corresponding to complete steady-state inversion of the lasing levels.)

\subsection{General Solution of the SPA-SALT Equations}

Let us rewrite the SPA-SALT equation (\ref{eq:SPASALT}) as
\begin{equation}
\frac{D_0}{D_0^\mu} -1 =  \sum_\nu A_{\mu\nu} \, I_\nu, \qquad
A_{\mu\nu} \equiv \Gamma_\nu \, \chi_{\mu \nu}. \label{eq:SPASALT2}
\end{equation}
This seems to be simply an inhomogeneous linear system to be solved by inversion, but in fact it is more complicated, for we have not indicated the number of modes to be summed over.  Let us suppose that we have solved the non-interacting threshold conditions (\ref{threshold cond 1}) and (\ref{threshold cond 2}) for a given $\epsilon_c(\vec{r})$ and $F(\vec{r})$, obtaining a subset of $M$ TLMs $\{u_\mu(\vec{r}) \, |\, \mu = 1, 2, \cdots M\}$, with real non-interacting thresholds $D^\mu_0$ less than some cut-off value, $D_{0,c}$ (taken to be much higher than the first lasing threshold).   For a given $D_0$, the indices $\mu, \nu$ occurring in (\ref{eq:SPASALT2}) are those corresponding to lasing modes that have turned on.  We have used this fact in deriving (\ref{eq:SPASALT}), where we divided out the common factor $a_\mu$, which is valid only if $a_\mu$ is non-zero.  Hence (\ref{eq:SPASALT}) is a {\it constrained} inversion problem; for each value of $D_0$, we must construct the matrix $A_{\mu \nu}$ from the correct subset of the $M$ TLMs at our disposal.

We wish to find an ordered set of matrices $A^{(1)}_{\mu \nu}, A^{(2)}_{\mu \nu}, \ldots A^{(N_{\textrm{max}})}_{\mu \nu}$,
as well as the associated \textit{interacting} thresholds $D_{0,\textrm{int}}^\mu$, which are the values of $D_0$ at which the $\mu$-th mode turns on.  Because the SPA-SALT includes the effects of nonlinear modal interactions, these differ from the non-interacting thresholds $D_0^\mu$.  In fact, $N_{\textrm{max}}$ often is less than $M$, since some of the candidate modes may never turn on at any pump value, as we will see below.  For a given $D_0$, let us suppose that $N$ lasing modes have turned on.  Without loss of generality, we assume that the indices for these lasing modes are $\mu = 1, \cdots, N$.  We now have a non-sparse $N\times N$ matrix $A_{\mu \nu}$, and can invert (\ref{eq:SPASALT}) to obtain
\begin{align}
  \begin{aligned}
    I_\mu &= c_\mu D_0 - b_\mu, \quad & \mu &= 1,2,\ldots N \\
    c_\mu &= \sum_{\nu=1}^N \frac{(A^{-1})_{\mu\nu}}{D_0^\nu}, \quad &
    b_\mu &= \sum_{\nu=1}^N (A^{-1})_{\mu\nu}. \label{eq:linear}
  \end{aligned}
\end{align}
From this, we see that the intensity of each lasing mode increases linearly with $D_0$, between each threshold, no matter how many modes are lasing or how far the laser is above threshold.

To find the next matrix $A^{(N+1)}_{\mu \nu}$ we must find the lowest interacting threshold $D_{0,\textrm{int}}^{N+1}$ for the remaining set of $M-N$ modes.  To do this, we note that (\ref{eq:linear}) is valid for $D_{0,\textrm{int}}^N \le D_0 \le D_{0,\textrm{int}}^{N+1}$ (the lasing intensities are continuous at each threshold although their slopes are not).  At the upper limit of this range, $D_0 = D_{0,\textrm{int}}^{N+1}$, we can equally well add mode $(N+1)$ to this matrix equation.  The resulting equation would yield identical solutions for $I_1,\cdots, I_N$, plus the solution $I_{N+1} = 0$.
Thus we can evaluate (\ref{eq:SPASALT2}) for all choices $\mu = N+1,\ldots M$:
\begin{equation}
  D_{0,\textrm{int}}^{\mu} = D_0^{\mu} \left[ 1 + \sum_{\nu = 1}^N A_{\mu \nu}(c_\nu D_{0,\textrm{int}}^{\mu} - b_\nu) \right],
\end{equation}
which gives $N-M$ explicit linear relations for the possible $N+1^{st}$ threshold.  Evaluating these relations, one simply chooses the lowest value, which is then the correct $N+1^{st}$ interacting threshold. This defines a recursive procedure to find all the interacting thresholds and uniquely determine the ordered set of $A$ matrices required to compute $I_\mu(D_0)$ for the entire desired range of $D_0$.

Note that we always assume the ``non-trivial zero'' solution at each (interacting) threshold, i.e.~that the physical solution switches from the trivial zero for $I_{N+1}$ to the non-zero lasing solution, giving rise to a bifurcation with discontinuous slope.  When this happens, all the modes which are already turned on experience a negative kink in their slopes at higher thresholds.  This behavior is characteristic of lasers when higher modes turn on, and the SPA-SALT captures it in a simple manner.

Once the constraints on Eq.~(\ref{eq:SPASALT2}) are implemented in this manner, the solution of the SPA-SALT equations requires just $N_{max}$ inversions of relatively small matrices generated from the input parameters,
$\{\chi_{\mu\nu}\}, \{\Gamma_\nu\},\{D_0^\nu\}$.
{\it Thus the computational time for solving the SPA-SALT equations is negligible once the TLMs have been calculated.} When the single-pole approximation is good, the nonlinear multimode problem becomes only minimally harder than the linear TLM problem, which can be adapted for efficient solution using finite element or boundary element methods \cite{Wiersig_BEM,Chong}.  In Section~\ref{sec:test_SPASALT} we compare the SPA-SALT lasing solutions to the exact SALT calculations, finding good agreement.  Note that it has already been shown \cite{TSG OE} that the exact SALT solutions agree to within a few percent with exact time-dependent MB simulations for simple 1D edge-emitting lasers, as long as the conditions for the stationary inversion approximation are well-satisfied.

\subsection{Gain-clamping Transition}

Eq.~(\ref{eq:linear}) gives a linear relation determining each of the $N_{max}$ interacting thresholds of the form
\begin{equation}
 D_{0,\textrm{int}}^{\mu} = f_\mu(\{\chi_{\mu \nu}\}, \{\Gamma_\nu\},\{D_0^\nu\}) \,D_0^\mu \equiv
 \frac{1}{1-\lambda_\mu} D_0^{\mu},
 \label{eq:thresholds}
\end{equation}
where the function $f_\mu \equiv (1-\lambda_\mu)^{-1}$, is the threshold enhancement factor which increases the $\mu^{\textrm{th}}$ threshold from its non-interacting value, due to the spatial hole-burning of lower threshold modes, which depletes the gain.  In simplified treatments of the laser rate equations, in which the cavity mode is assumed perfectly uniform in space, these interactions actually clamp the effective gain so that it no longer increases with the external pump, predicting that no additional modes turn on \cite{Haken}.  In reality all resonators admit multiple modes with incomplete spatial overlap and so this extreme gain clamping behavior is not realized.  The SPA-SALT gives a much more rigorous criterion for gain-clamping at the level of the $N^{\textrm{th}}$ lasing mode.  If $\lambda_N \to 1$ then all higher thresholds are pushed off to infinity and no more modes can turn on for any value of the pump.

Note the analogy here to mean-field phase transitions, for example where a strong enough magnetic interaction causes the susceptibility to diverge.  Here strong interactions, meaning large values of the coefficients $\chi_{\mu \nu} (\mu \neq \nu)$, suppress ``ordering" of higher modes.  Conversely, spatially disjoint or weakly overlapping modes will not be suppressed and their interacting threshold will be approximately equal to their non-interacting thresholds.  In addition, higher modes with substantially lower modal gain and Q-values with respect to the first mode(s) will be more easily suppressed.  Calculations for various examples indicate that this gain-clamping ``phase" of the laser can be reached for realistic lasers.  We calculate and discuss the coefficient $\lambda_2$ below.

\subsection{One- and Two-mode Solutions}

To get a feeling for the SPA-SALT solutions, we now present explicit results for one and two mode lasing, which illustrate most of the qualitative features of the theory.  The single mode result is trivial.  The lowest non-interacting threshold, $D_0^{(1)}$, is found as part of the calculation of the initial set of $N$ TLMs, and of course is the correct first threshold.  Eq.~(\ref{eq:SPASALT2}) is just a scalar equation for the first mode intensity, yielding
\be
I_1 = \frac{1}{\Gamma_1\chi_{11} D_0^{(1)}} (D_0-D_0^{(1)}) \label{eq:singlePole_1mode},
\ee
where $\chi_{11}^{-1} \equiv V_1$ plays the role of the mode volume, enhancing the power slope if mode one is more evenly distributed over the gain volume. We should point out that $I_1$ should be thought of as the intensity within the cavity. The emitted power is found by integrating the photon flux associated with the TLM $u_\mu (\vec{r})$ over a surface at infinity \cite{TS}; the transmissivity of the cavity is implicitly contained in the calculation of the TLM.  In Appendix A we show that this power output can be related to a volume integral of the TLM over the gain region of the cavity, and that for single-mode lasing within the SPA-SALT one finds

\begin{equation}
  \mathcal{P}_1 = \frac{k_1}{2\pi}\, \frac{\int\! d^dr F(\vec{r})
    |u_1|^2}{\int\! d^dr F(\vec{r}) u_1^2 |u_1|^2} \left(D_0 - D_0^{(1)}\right).
\end{equation}
Recently this equation was found to agree very well with the output power of a novel surface-emitting photonic crystal laser calculated using non-linear FDTD methods \cite{pcsel}.

Using the procedure described above we now determine the mode $\nu$ with the lowest interacting threshold and
the correct $2 \times 2$ matrix
$A^{(2)}_{\mu \nu}$ above this threshold.  The second threshold is found to be
\be
D_{0,\textrm{int}}^{(2)} = \frac{\chi_{11}-\chi_{21}}{\chi_{11}-\frac{D_0^{(2)}}{D_{0}^{(1)}}\chi_{21}}\,D_{0}^{(2)} \equiv \frac{1}{1-\lambda_2} D_0^{(2)} \label{eq:2ndTH}
\ee
where the interaction coefficient
\be
\lambda_2 = \left[\frac{D_0^{(2)}}{D_0^{(1)}}-1\right]\,\frac{\chi_{21}}{\chi_{11} - \chi_{21}} \geq 0.
\ee
Note that, as $D_0^{(2)} > D_0^{(1)}$, as long as the modal interaction coefficient $\chi_{21}$ is non-vanishing, the
interacting second threshold is higher than the non-interacting threshold.
The gain clamping limit is reached when $\lambda_2 \to 1 \Rightarrow \chi_{21} \to  \chi_{11}D_0^{(1)}/D_0^{(2)} $, and the first mode suppresses any second mode for all values of the pump.  One sees that strong overlap $\chi_{21} \approx
\chi_{11}$ leads to gain clamping as we expect.  Also if the second mode has significantly lower Q-value or is away
from the center of the gain curve, the ratio $D_0^{(1)}/D_0^{(2)}$ is reduced leading to gain clamping for smaller values
of $\chi_{21}$. One way to achieve this limit is in a microcavity laser with passive cavity modes spaced more widely than the gain bandwidth.

When the pump exceeds the second threshold $D_{0,\textrm{int}}^{(2)}$, the modal intensities $I_1$ and $I_2$ are obtained
from Eq.~(\ref{eq:SPASALT2}),
\bea
 I_1 &=& \frac{\chi_{22}/D_0^{(1)}- \chi_{12}/D_0^{(2)}}{\Gamma_1(\chi_{11}\chi_{22}-\chi_{12}\chi_{21})}\,(D_0-D_0^{\prime (1)}), \label{eq:int1_TLM2m1p}\\
 I_2 &=& \frac{\chi_{11}/D_0^{(2)}- \chi_{21}/D_0^{(1)}}{\Gamma_2(\chi_{11}\chi_{22}-\chi_{12}\chi_{21})}\,(D_0-D_{0,\textrm{int}}^{(2)}). \label{eq:int2_TLM2m1p}
\eea
where the modified intercept $D_0^{\prime (1)}$ is given by
\be
D_0^{\prime (1)} = \frac{\chi_{22}-\chi_{12}}{\chi_{22}-\frac{D_0^{(2)}}{D_0^{(1)}}\chi_{12}}  D_{0,\textrm{int}}^{(1)}.
\ee
The change in intercept indicates that the first mode intensity has a negative kink at the second mode threshold
($D_0^{\prime (1)}< D_{0,\textrm{int}}^{(1)}$),
as can also be seen directly from the slope of $I_1$, which is reduced from its value of $1/(\Gamma_1 \chi_{11} D_0^{(1)})$ in the interval below the second threshold. This kink is always negative because the turning on of a second mode reduces the slope efficiency of the laser in the first mode, but vanishes
when the interaction coefficient $\chi_{12} \to 0$ and the two lasing modes act independently.

\subsection{Tests of the SPA-SALT}
\label{sec:test_SPASALT}

\begin{figure}
\centering
\includegraphics[width=7cm]{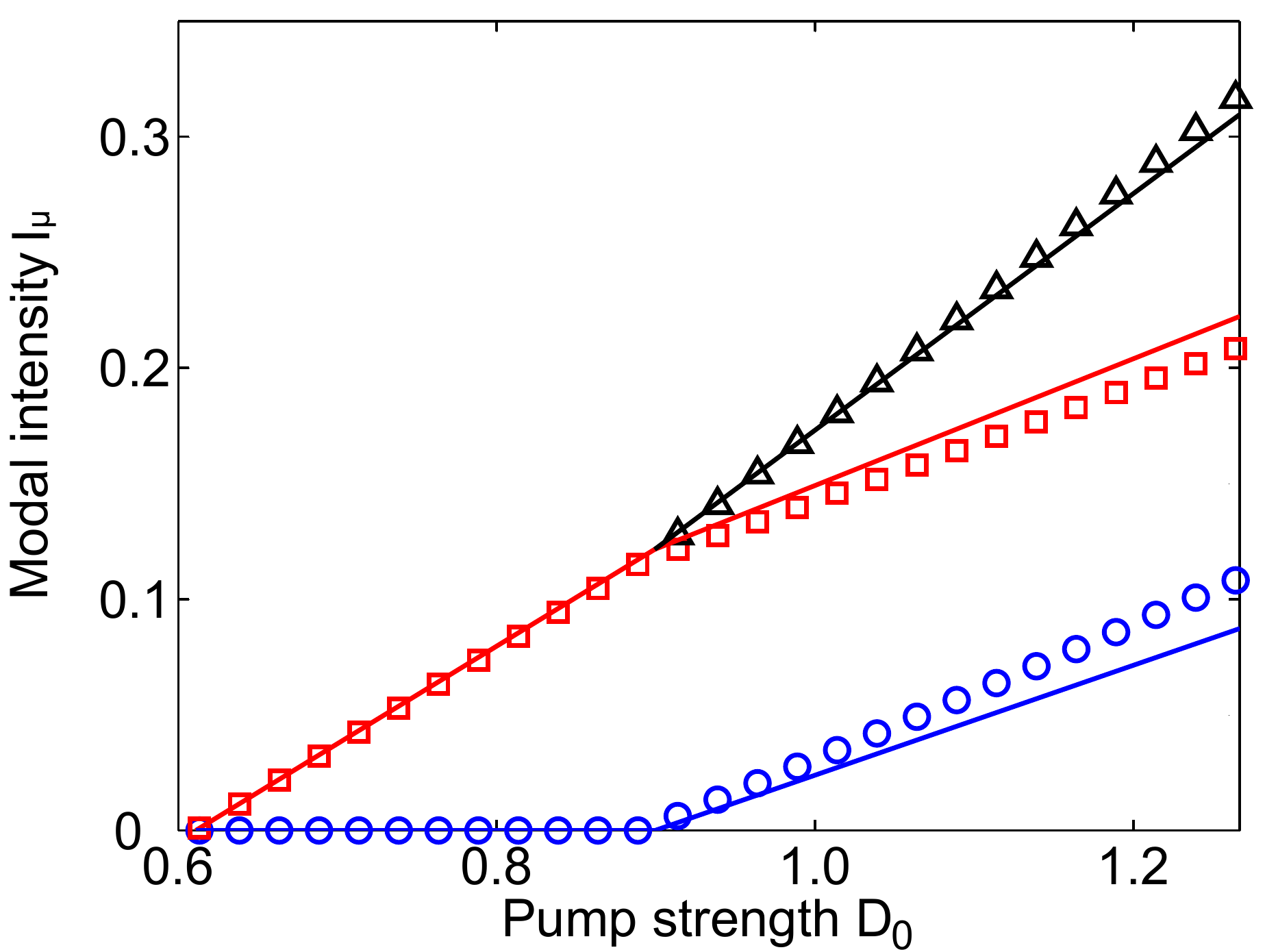}
\caption{(Color online) Modal intensity versus pump strength in a 1D slab resonator. The description of the resonator is given in the caption of Fig.~\ref{fig:CFvsMLG1d}. Open symbols show the numerical solutions of (\ref{eq:SALT_TCF}) and solid lines are the results of single-pole approximation (Eqs.~(\ref{eq:singlePole_1mode},\ref{eq:int1_TLM2m1p}-\ref{eq:int2_TLM2m1p})). The color scheme is: blue (Mode 1), red (Mode 2), black (total intensity in the two-mode regime). }
\label{fig:singlePole}
\end{figure}

To test the results derived above, we first revisit the 1D laser studied in Section~\ref{threshold discussion}. Fig.~\ref{fig:singlePole} shows the growth of modal intensities with $D_0$.  In the single-mode regime, the result given by (\ref{eq:singlePole_1mode}) agrees very well with the numerical solution of (\ref{eq:SALT_TCF}), indicating that the single-pole approximation is almost exact.  Consequently, the second threshold $D^{(2)}_0 = 0.892$ is also accurately predicted by (\ref{eq:2ndTH}), which gives $D^{(2)}_0 = 0.899$. In the two-mode regime we still find good agreement, but the SPA-SALT slightly overestimates the suppression of the second mode. Nevertheless, the total intensity is in good agreement with the full SALT solution.

\begin{figure}
\centering
\includegraphics[width=7cm]{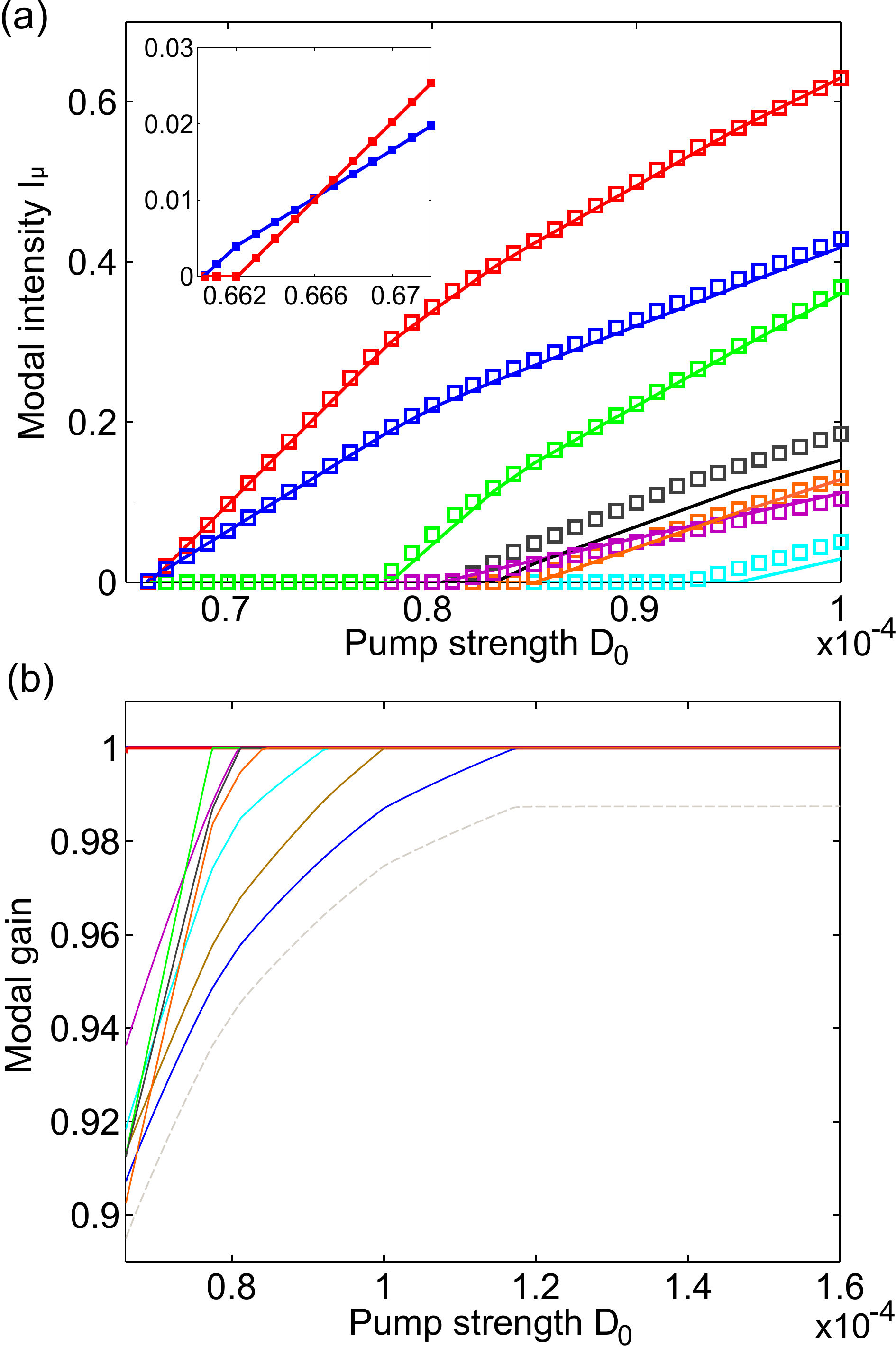}
\caption{(Color online) (a) Modal intensity versus pump strength in a 2D disk laser of uniform index
$n=3.3 + 10^{-5}i$, uniformly pumped. Squares show the numerical solutions of (\ref{eq:SALT_TCF}) and solid lines are the results of the single-pole approximation (\ref{eq:linear}). Inset: Zoomed view near the first two thresholds. (b) Modal gain versus pump strength for the first ten TLMs, calculated with the full SALT, indicating that mode ten will never turn on due to modal interactions, as predicted by the SPA-SALT. The dashed line indicates the fully suppressed tenth mode. The first two modes are too close together to be distinguished in this plot. Modal gain is defined in terms of eigenvalues of the
modified lasing map and a mode reaches threshold when the modal gain reaches unity \cite{TSG Nonlinearity}.}
\label{fig:singlePole_disk}
\end{figure}

To demonstrate the accuracy of the SPA-SALT in cases where the mode density is high, we study a uniformly pumped 2D disk laser of radius $R=1$ and index $n=3.3+10^{-5}i$. The gain is assumed to center at Re$[nk_aR]=66$ with width $\gper = k_a/40$. Now there exist high-Q whispering gallery modes, and we find that first two thresholds are very closely spaced (see inset; Fig.~\ref{fig:singlePole_disk}) and are four orders of magnitude smaller than those in the 1D example just treated. The SPA-SALT correctly captures the intensity crossover of the first two modes shortly after the second one turns on, and its prediction for the first three modes remains impressively accurate, even after the onset of the 7th mode. As we have seen in Fig.~\ref{fig:aboveTH}(b), the higher order mode(s) are less single-pole like compared to the lower order ones. Thus we expect the SPA-SALT not to work as well for higher order modes; this can be seen from the noticeable differences in the 5th (black) and 7th (cyan) thresholds given by the SPA-SALT and the full SALT results. Nevertheless, the slopes of all the higher-order modal intensities are still largely correct.

As noted, Eq.~(\ref{eq:thresholds}) gives a criterion for a complete suppression of modes after a certain number of modes $N$ have turned on.  Typically if a mode is completely suppressed this equation gives a negative (unphysical) result.  This happens for the tenth TLM in the current example.  Indeed, the full SALT calculation, using the modified threshold matrix \cite{TSG Nonlinearity}, confirms the prediction that mode ten will never turn on (see Fig.~\ref{fig:singlePole_disk}(b)).

\begin{figure}
\begin{center}
\includegraphics[width=7cm]{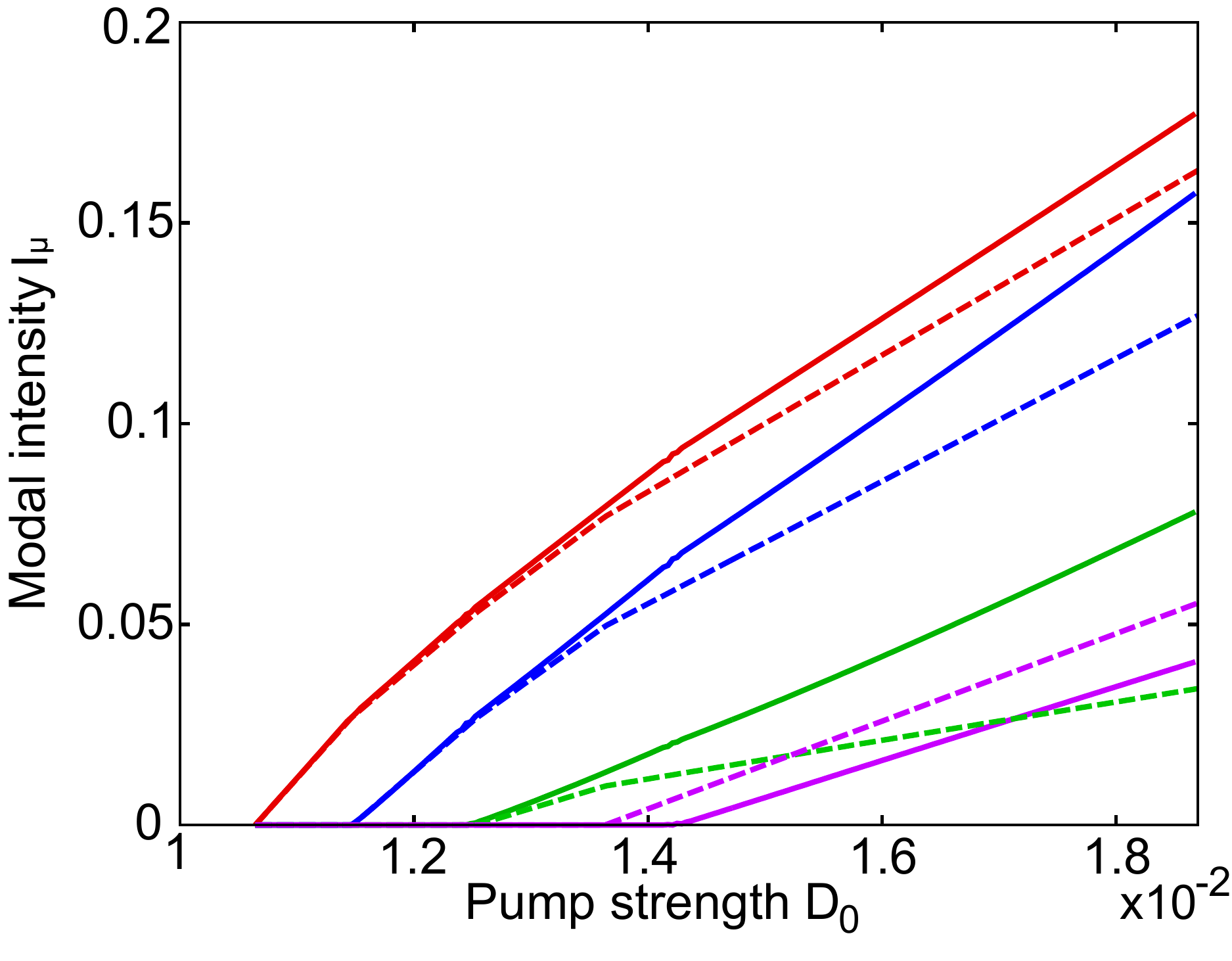}
\caption{(Color online) Modal intensity versus pump strength in a random laser.  The cavity is a disk of radius $R$, uniformly pumped with $k_a = 60/R$ and $\gamma_\perp = 1/R$.  The scattering mean free path is $\ell = R/3$, significantly shorter than the systems studied in Ref.~\cite{TSG Science}.  The solid curves give the exact SALT solution, and the dashed curves the SPA-SALT solutions. }
\label{fig:SPA_breakdown}
\end{center}
\end{figure}

The SPA is better satisfied the less open is the laser cavity.  The random laser is a system in which there is no conventional cavity, only multiple scattering to slow escape.  In the most challenging case of a weakly scattering RL, it has no sharp linear resonances at all, only the presence of the gain medium allows strong preference for certain frequencies \cite{TSG Science}.  In Ref.~\cite{TSG Science} the modal intensities for a 2D RL were found within the full SALT theory to be a nonlinear function of the pump, unlike all other cases studied.  Thus we do not expect the SALT to apply there.  Even when the disorder scattering is increased in the RL in order to increase the Q, and the intensities are linear in the pump, we find that the SPA-SALT, while it still
gives good qualitative results, does not give good quantitative agreement with the exact SALT solutions, as shown in Fig.~\ref{fig:SPA_breakdown}.

\section{Summary and conclusions}

We have presented an improvement of steady-state ab initio laser theory based on the TCF basis that allows one to solve the self-consistent SALT equations more efficiently for resonators which are spatially inhomogeneous (as is usually the case) and/or with inhomogeneous pumping. This completes the development of the ab-initio theory based on the stationary inversion approximation, originally proposed in 2006 \cite{TS} and improved in several subsequent papers \cite{TSG PRA,TSG OE,TSG Science,TSG Nonlinearity}.  This theory takes into account the openness of the cavity exactly in terms of TCF or UCF basis states, and includes the nonlinear hole-burning interactions to infinite order.  Besides predicting interacting thresholds and intensities, the theory captures subtle effects such as the change in shape of the lasing modes, and the variations in their frequencies as the pump is increased well above threshold.

Using the TCF basis and the single-pole approximation \cite{TS}, we have derived a simplified version of the theory, the SPA-SALT, which predicts a linear increase of all lasing intensities.  The relevant slopes and interacting thresholds can be found with negligible computational effort, once the linear problem of the non-interacting threshold lasing modes is solved.  Explicit analytic solutions were for the few-mode lasing regime, illustrating important qualitative features of multimode lasing with modal interactions.  In particular, an analytic condition was found for the ``gain-clamping"
transition, in which higher modes are completely suppressed by modal interactions.  For non-trivial examples, the SPA-SALT agrees well with the full SALT, although its breakdown for very low-Q systems such as random lasers was also found.  Although further work is needed to determine the regime of quantitative validity of the SPA-SALT, there is already evidence that it will be possible to dramatically improve the modeling of realistic and complex laser structures in two and three dimensions  \cite{pcsel}, by reducing the nonlinear lasing computation to almost the same level of difficulty as the linear problem of finding the threshold lasing modes.
Even when the SPA-SALT is not a good approximate theory, the full SALT equations in the TCF basis will improve steady-state lasing calculations by many orders of magnitude compared to brute force time-domain simulations.

\section{Acknowledgments}

This work was partially supported by NSF Grants No. DMR-0808937 and No. DMR-0908437, seed funding from the Yale NSF-MRSEC (DMR-0520495), and by the facilities and staff of the Yale University Faculty of Arts and Sciences High Performance Computing Center.  We acknowledge helpful discussions with Hui Cao.

\appendix
\section{Modal output power}
\label{sec:outputpower}

In this appendix we derive the modal output power of a 2D cavity from the (internal) modal intensity. The output power is the total flux of the Poynting vector, taken across a loop $\Gamma$ enclosing the cavity:
\begin{equation}
  \mathcal{P} = \frac{1}{4\pi} \oint_\Gamma ds \, \hat{n} \cdot \left[\vec{E} \times \vec{B}\right].
\end{equation}
In (\ref{mode ansatz}), the (out-of-plane) electric field is written as a sum over the modal fields $\Psi_\mu(\vec{r})$, and a similar expression may be written for the (in-plane) magnetic field. We find the time-averaged total output power $\langle\mathcal{P}\rangle = \sum_\mu \mathcal{P}_\mu$, with the modal power $\mathcal{P}_\mu$ given by
\begin{eqnarray}
  \mathcal{P}_\mu &=& \frac{i}{4\pi k_\mu} \oint_\Gamma\! ds\;
\hat{n} \cdot \left[\Psi_\mu \nabla \Psi_\mu^* - \textrm{c.c.}\right] \\
   &=& \frac{i}{4\pi k_\mu} \int_C\! d^2r\;
\left[\Psi_\mu \nabla^2 \Psi_\mu^* - \textrm{c.c.}\right].
  \label{output power integral}
\end{eqnarray}
In the last step we have used the Gauss' law. Here $\Psi_\mu$ and $\mathcal{P}_\mu$ are measured in their natural units $e_c$ and $e_c^2$, introduced when deriving Eq.~(\ref{TSG2}). Using Eq.~(\ref{output power integral}), together with the wave equation (\ref{TSG1a}) and its
complex conjugate, gives us Eq.~(\ref{outputpower}), which we
reproduce here for convenience:
\begin{equation}
  \mathcal{P}_\mu = \frac{k_\mu}{2\pi} \, \int_C d^2r\,
\left\{\frac{\Gamma_\mu D_0 F(\vec{r})}{1 + h(\vec{r})} -
\textrm{Im}[\epsilon(\vec{r})]\right\} \, |\Psi_\mu(\vec{r})|^2.
\end{equation}
This result states that the total power radiated by each lasing mode equals
the power that the gain medium delivers into that mode, minus the
power that the mode loses through material dissipation (described by
$\textrm{Im}[\epsilon]$).

It is instructive to consider the modal power in the single-pole
approximation.  Let us suppose that $\textrm{Im}[\epsilon] = 0$.
Combining the general expression for $\mathcal{P}_\mu$ in (\ref{output
power integral}) with the SPA-SALT expression $\Psi_\mu \approx
\sqrt{I_\mu} \, u_\mu$, we obtain
\begin{align}
  \mathcal{P}_\mu &= \frac{iI_\mu}{4\pi k_\mu} \int\! d^2r \left[u_\mu
    \nabla^2 u_\mu^* - \textrm{c.c.}\right] \\ &= \frac{k_\mu}{2\pi}\,
  \Gamma_\mu D_0^\mu I_\mu \int\! d^2r F(\vec{r}) |u_\mu|^2.
\end{align}
Here we have used (\ref{threshold_condition}) to express
$\textrm{Im}[\eta_\mu]$ in terms of the SPA laser threshold $D_0^\mu$.
As noted in the main text, in the single-mode regime ($\mu = 1$), the modal power has a
particularly simple form: using (\ref{eq:SPASALT}), we can write
$I_\mu$ in terms of the pump $D_0$, to obtain
\begin{equation}
  \mathcal{P}_1 = \frac{k_1}{2\pi}\, \frac{\int\! d^2r F(\vec{r})
    |u_1|^2}{\int\! d^2r F(\vec{r}) u_1^2 |u_1|^2} \left(D_0 - D_0^{(1)}\right).
\end{equation}

\section{Comparison to Mandel approach}

In Refs.~\cite{Mandel1,Mandel2} Mandel and coworkers treated the infinite-order modal interactions in a Fabry-Perot cavity in the single-mode and two-mode regimes. They used the approximations of stationary inversion, and pump-independent lasing modes and frequencies, similar to the SPA-SALT (the SALT of course includes the pump-dependence of the lasing modes and frequencies \cite{TSG Science}).  Unlike the SPA-SALT, they assumed that the fixed lasing modes were hermitian closed cavity modes (sine waves of real wavevector).  They did not derive a version of the basic constrained linear equation (\ref{eq:SPASALT}) of the SPA-SALT, but instead they derived a single-pole closed-cavity version of Eq.~(\ref{eq:SALT_TCF}).  For the single-mode case, Eq.~(5) of Ref. \cite{Mandel1} is of exactly the same form as Eqs. (43),(54) of Ref.~\cite{TS}, the earliest version of the SALT, except for their use of closed cavity modes.  Ref.~\cite{TS} applies the single-pole approximation to the direct map but treats the openness of the cavity exactly using non-hermitian constant flux states; this approximation is not exactly equivalent to the SPA-SALT, which uses the SPA on the inverse map, but gives very similar results to the SPA-SALT at large pump strength.

It is interesting to compare the two methods for the simple case of a uniformly pumped 1D dielectric slab laser of the type considered in \cite{TS,TSG PRA,TSG OE} (see inset, Fig.~\ref{fig:smatPoles}). We will compare Mandel's approach to the full SALT, the most complete form of our theory. Thus our approach differs from Mandel in two major ways. First we take into account the openness of the cavity exactly and second we allow for the change in the lasing modes and modal frequencies above threshold. To vary the quality factor of the cavity, we choose four sets of parameters $\{n,k_a\}=\{1.5,40\},\{3,20\},\{5,20\}$, and $\{10,20\}$ ($k_a$ is the frequency of the gain center). We have shown in Ref.~\cite{TSG OE} that for the first two sets of parameters the SALT and numerical solutions of the MB equations agree very well, so we can take the SALT results as correct.

The rescaled model intensity ($I'_\mu \equiv \Gamma_\mu I_\mu$) in the single-mode regime in Mandel's approach is given in our notation by
 \be
  I'(D_0) = \frac{1}{4}\left(4\frac{D_0}{D_0^{(1)}} - 1 - \sqrt{8\frac{D_0}{D_0^{(1)}}+1}\right). \label{eq:Mandel1}
 \ee
The dependence on the refractive index of the cavity is contained in the first threshold, $D_0^{(1)}$, which is not calculated in the Mandel approach, but is assumed known and used to normalize the pump. The gain parameters ($k_a$ and $\gamma_\perp$) only enter in the scale factor ($\Gamma_\mu$) and implicitly again through  $D_0^{(1)}$.  Note that the Mandel single-mode result has an additional square root dependence on the pump, which is not present in the SPA-SALT.  This difference arises because, as already noted, the single-pole approximation is made at a different point in the two derivations.  The full SALT theory does not predict a universal linear dependence on pump and indeed for very low-Q lasers, such as random lasers, the dependence can be non-monotonic \cite{TSG Science}.

\begin{figure}[htbp]
\centering
\includegraphics[width=7cm]{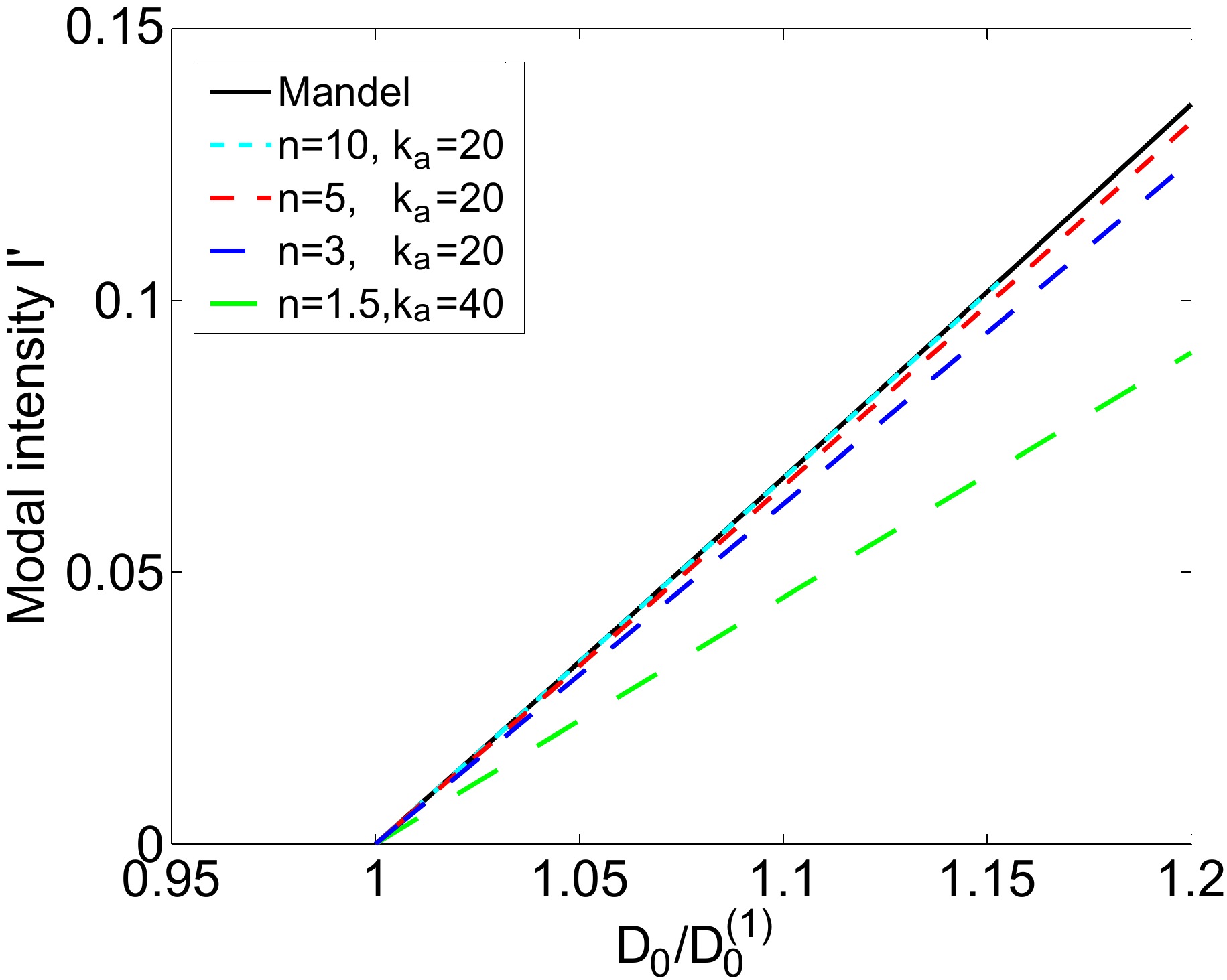}
\caption{(Color online) Rescaled intensity $I_\mu' \equiv \Gamma_\mu I_\mu$ of the first mode in a 1D slab resonator. The cavity is open on both sides and the pump is taken to be spatially uniform. $D_0$ is the pump intensity and $D_0^{(1)}$ is its threshold value. The solid line is produced using Eq.~(\ref{eq:Mandel1}), which has no dependency on the cavity index or length. The four dashed lines are the results of the SALT with different cavity indices and atomic transition frequecies. The upper bound of single-mode lasing in the $n=10$ (high-Q) case is near $D_0/D_0^{th}\approx1.15$, and the intensity overlaps with the solid curve. As the cavity index/Q-factor is reduced, the black curve differs more and more from the SALT result, whose accuracy has been proven by comparing with the time-dependent simulation of the steady-state solutions of the MB equations (Ref.~\cite{TSG OE}).}\label{fig:vsMandel1}
\end{figure}

In Fig.~\ref{fig:vsMandel1} we compare the result given by Eq.~(\ref{eq:Mandel1}) to the SALT. As one might have expected, the two approaches agree well for the higher Q cases ($n=10,5$) but a significant disagreement in the slope of the intensity curves appears for the lower Q ($n=1.5,3$) cases.  Nonetheless, the Mandel approach for the single-mode case is qualitatively better than HS, which shows an unphysical saturation \cite{TS,TSG OE}.

\begin{figure}[htbp]
\centering
\includegraphics[width=7cm]{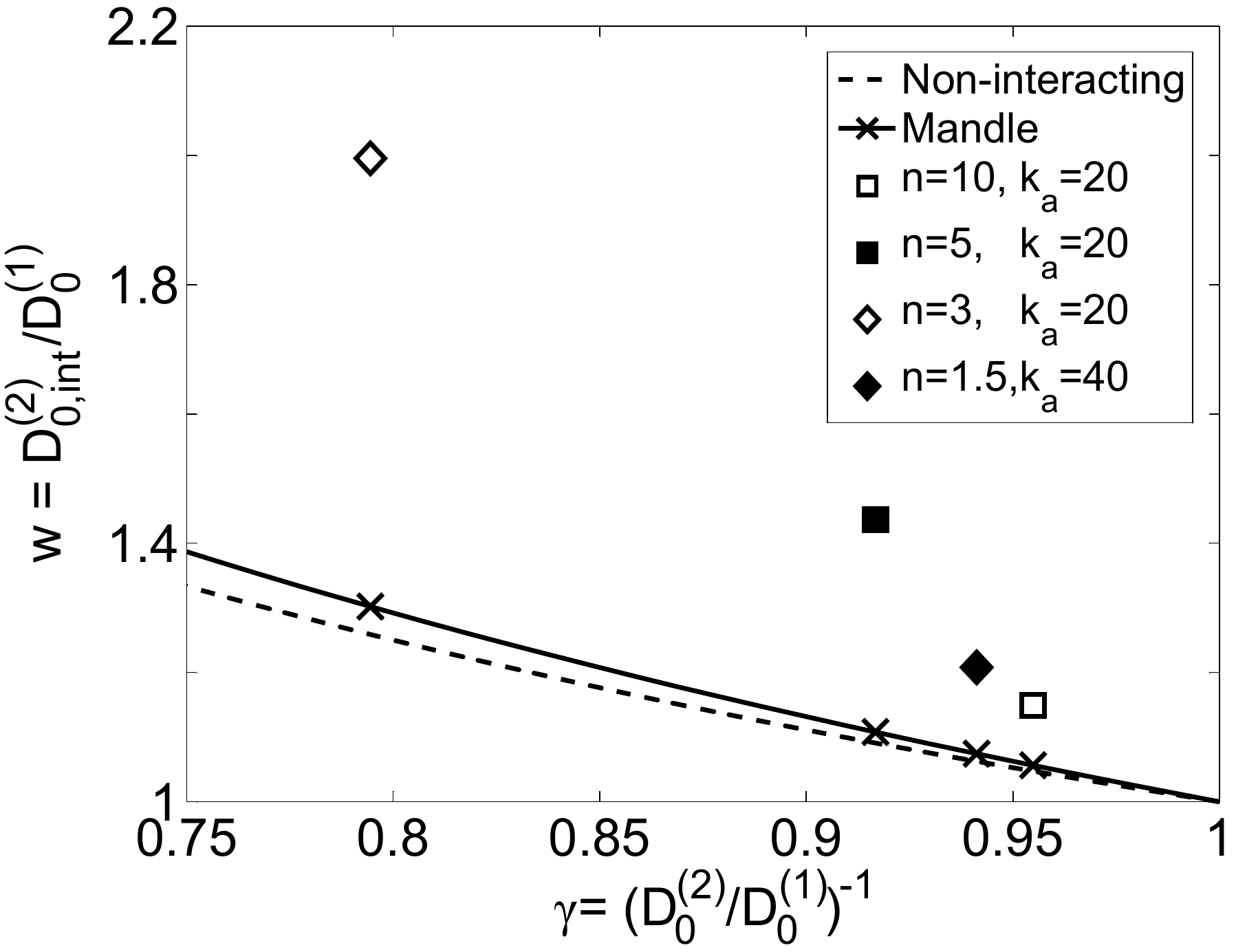}
\caption{Second interacting threshold in a 1D slab resonator. The cavity and parameters used are the same as in Fig.~\ref{fig:vsMandel1}. $D_0^{(2)}$ is the second threshold value in the absence of modal interaction. The solid line and crosses are the solution of Eq.~(\ref{eq:Mandel2}), and the dotted line indicates the non-interacting case ($\gamma = 1/w$). The results of the SALT are indicated by the different symbols explained in the legend.}
\label{fig:vsMandel2}
\end{figure}

Next we compare the value of the interacting second threshold $D_{0, \text{int}}^{(2)}$ given implicitly in Mandel's method by
\be
I'(D_{0,\text{int}}^{(2)})\left(\frac{D_0^{(1)}}{D_0^{(2)}} + 2 - 2\frac{D_{0,\text{int}}^{(2)}}{D_0^{(2)}}\right)^2
= 4\frac{D_0^{(1)}}{D_0^{(2)}}\left(\frac{D_{0,\text{int}}^{(2)}}{D_0^{(2)}}-1\right) \label{eq:Mandel2}
\ee
and the result of the SALT in the four cases listed above. We find that Mandel's approach consistently underestimates the strength of the modal interactions and deviates relatively little from the non-interacting threshold values (see Fig.~\ref{fig:vsMandel2}).  The highest Q case agrees most closely with the SALT, but there is some non-monotonic behavior of the thresholds with Q value in the SALT which we did not analyze in detail. We conclude that the effect of openness accounts for the main difference between the SALT and the Mandel approach, in a Fabry-Perot cavity in which both can be applied.  Mandel's approach is qualitatively better than that of HS but is not as accurate as the SALT and the SPA-SALT, both of which are based on general computational algorithms applicable to arbitrary cavities.

\section{Perturbative calculation of corrections to the SPA-SALT}

\begin{figure}[htbp]
  \centering
  \includegraphics[width=7cm]{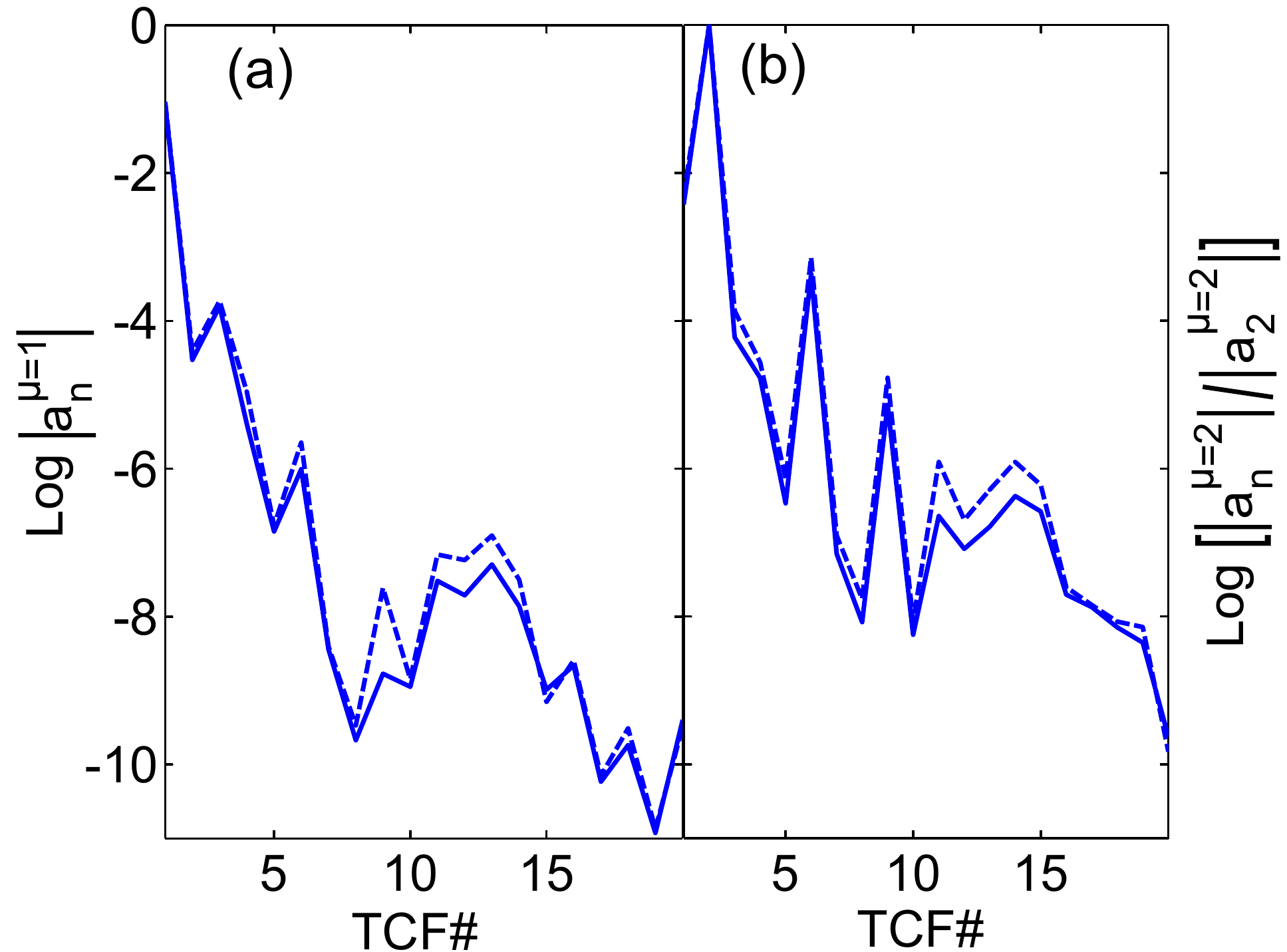}
  \caption{Expansion coefficients of the first lasing mode (left) and the second lasing mode (right) at the second threshold in a 1D slab resonator. The solid curve is the solution of (\ref{eq:SALT_TCF}) and the dashed line is given by the approximation (\ref{eq:multiCorr_sideband}) and (\ref{eq:multiCorrMode2_sideband}), respectively. Notice that the vertical axis is in the logarithmic scale; the expansion of the first/second mode is dominated by the first/second UCF state with a weight of $90\%/84\%$.} \label{fig:multipoleCorr2}
\end{figure}

The major approximation in the SPA-SALT is replacing the expansion (\ref{eq:TCFexpansion}) with a single term,
$\Psi_\mu = a^\mu_1 u_\mu \equiv a_\mu u_\mu$.
In this appendix we derive the first order expression for the non-dominant expansion coefficients $a^\mu_m$ in the single-mode regime. Assuming the dominant component is $a_1$, we approximate $h(\vec{r})$ by $|a_1 u_1(\vec{r})|^2 = I_{1}|u_1(\vec{r})|^2$. Eq.~(\ref{eq:inverse}) for $a_{n(n>1)}$ is then
\bea
&&D_0 \,a_{n} = \frac{a_{n}}{\lambda_{n}} + \Gamma_1I_1\sum_m \frac{\chi_{nm1}^{(1,1)}}{\lambda_m}\,a_m,  \label{eq:multipoleCorr}\\
&&\chi_{nmn'}^{(\mu,\nu)} \equiv \int d^dr F(\vec{r}){u_n(\vec{r},k_\mu^{(t)})}{u_m(\vec{r},k_\mu^{(t)})}|u_{n'}(\vec{r},k_\nu^{(t)})|^2. \nonumber
\eea
By inserting the expression (\ref{eq:singlePole_1mode}) for $I_1$, derived in the single-pole approximation, into Eq.~(\ref{eq:multipoleCorr}), we reduce the latter to a set of inhomogeneous linear equations of $a_{n(n>1)}$.
Eq.~(\ref{eq:multipoleCorr}) can be further simplified by keeping only the $a_1$ term in the sum, which leads to
\be
a_n = \frac{\chi_{n11}^{(1,1)}}{\chi_{111}^{(1,1)}}\frac{D_0-D^{(1)}_0}{D_0-\frac{\eta_n(k_1^{(t)})}{\gamma_1}}\,a_1 \label{eq:multiCorr_sideband}
\ee
Note that $\eta_n(k_1^{(t)})/\gamma_1$ is not $D_0^{\mu=n}$, which is $\eta_{n=\mu}(k_\mu^{(t)})/\gamma_\mu$.
In Fig.~\ref{fig:multipoleCorr2}(a) we compare (\ref{eq:multiCorr_sideband}) to the numerical solution of (\ref{eq:SALT_TCF}), and they agree very well even in the logarithmic scale.  The system is the inhomogenous 1D resonator considered in the main text, and the pump is tuned to the second threshold ($D_0=0.892$).

We can also derive an analytical expression to evaluate the non-dominant expansion coefficients of the second mode when it turns on. We assume that its dominant component is $a^{\mu=2}_2$, and derive
\be
\frac{a^{\mu=2}_n}{a^{\mu=2}_2} = \frac{\chi_{n21}^{(2,1)}}{\chi_{111}^{(1,1)}}\frac{\frac{D^{(2)}_{0,\textrm{int}}}{D_0^{(1)}}-1}{\frac{D^{(2)}_{0,\textrm{int}}}{D_0^{(2)}}-\frac{\eta_n(k_2^{(t)})}{\gamma_2D_0^{(2)}}} \label{eq:multiCorrMode2_sideband}
\ee
in the same way (\ref{eq:multiCorr_sideband}) is derived. It is easy to check using (\ref{eq:SPASALT2}) that the ratio becomes 1 when $n=2$ as it should. The result above is compared with the multi-pole expansion (\ref{eq:SALT_TCF}) in Fig.~\ref{fig:multipoleCorr2}(b).


\begin{thebibliography}{99}
\bibitem{AISC} In earlier work we used a less memorable acronym, AISC (Ab Initio Self-Consistent laser theory), which we now have dropped.  We apologize for any confusion.
\bibitem{bibnote1} The implementation of the out-going boundary condition in various geometries can be found
in \cite{TS} and \cite{LiThesis}.
\bibitem{weightMethod} The weight used here is defined as $|\alpha^\mu_m|/\sum_n |\alpha^\mu_n|$.
\bibitem{intensity} Within the stationary inversion approximation and the SPA the phase of $a^\mu_n$ is meaningless, as we neglect phase relations between modes and there is only one coefficient in the expansions of $\Psi_\mu$.  Hence only the modulus $|a_\mu|\equiv|a^\mu_{n_0}|$ is meaningful, and we will see that the SPA-SALT solves only for $I_\mu = |a_\mu|^2$.
\end{thebibliography}
\end{document}